\documentclass{aa}  
\usepackage{graphicx}
\usepackage{txfonts}

\usepackage[colorlinks, citecolor=blue]{hyperref}
\usepackage{natbib}
\usepackage{color}

\begin{document} 

   \title{Galaxy groups within voids}


\author{M.~Argudo-Fernández\inst{1, 2} \and
G.~Torres-Ríos\inst{1} \and
P.~Vásquez-Bustos\inst{1} \and
S.~Verley\inst{1,2} \and
I.~Pérez\inst{1,2} \and
S.~Duarte Puertas\inst{1,2} \and
A.~Jiménez\inst{1} \and 
R.~García-Benito\inst{3} \and 
A.~Zurita\inst{1,2} \and
M.~Alcázar-Laynez\inst{1} \and 
B.~Bidaran\inst{1} \and
A.~Conrado\inst{3} \and
D.~Espada\inst{1,2} \and 
E.~Florido\inst{1,2} \and
R.~González Delgado\inst{3} \and
M.~Hernández-Sánchez\inst{4} \and 
I.~del Moral-Castro\inst{5} \and 
J.~Román\inst{6} \and
L.~Sánchez-Menguiano\inst{1,2} \and
S.~Subramanian\inst{7,8} \and
P.~Villalba-González\inst{9} 
}

\institute{Departamento de Física Teórica y del Cosmos, Edificio Mecenas, Campus Fuentenueva, Universidad de Granada, E-18071 Granada, Spain. \email{margudo@ugr.es}
\and
Instituto Carlos I de Física Teórica y Computacional, Facultad de Ciencias, Universidad de Granada, E-18071 Granada, Spain
\and
Instituto de Astrofísica de Andalucía (CSIC), PO Box 3004, 18008 Granada, Spain
\and 
Departament d'Astronomia i Astrofísica, Universitat de València, E-46100 Burjassot (València), Spain
\and
Instituto de Astrofísica, Facultad de Física, Pontificia Universidad Católica de Chile, Campus San Joaquín, Av. Vicuña Mackenna 4860, Macul, Santiago, Chile, 7820436 
\and
Departamento de Física, Universidad de Córdoba, Campus de Rabanales, Edificio Albert Einstein, E-14071 Córdoba, Spain
\and
Indian Institute of Astrophysics, Koramangala II Block, Bangalore-560034, India 
\and
Leibniz-Institut für Astrophysik Potsdam (AIP), An der Sternwarte 16, D-14482 Potsdam, Germany
\and
Department of Physics and Astronomy, University of British Columbia, Vancouver, BC V6T 1Z1, Canada
}

   \date{Received ; accepted }

 
  \abstract
   {Despite cosmic voids being vast and almost empty, galaxy aggregations do exist within them, although they are much sparser than in denser regions like walls, filaments and clusters. The internal matter distribution within voids might have an impact on the properties and evolution of void galaxies.}
   {In this work, we aim to identify and characterise a sample of galaxy groups within voids in the local Universe (z\,<\,0.08), taking into account the peculiarities of these vast and empty structures.}
   {The void galaxies used in this study are selected from a well-defined void galaxy sample, from which the parent sample of the Calar Alto Void Integral-field Treasury surveY (CAVITY) legacy project was drawn. To identify galaxy groups, we applied a fiends-of-friends (FoF) like group finder algorithm to the selected sample, ensuring a certain degree of gravitational binding among group members. The same algorithm has been applied to identify a control sample of groups not in clusters nor voids, referred as NCNV groups.} 
   {The catalogue of groups consists on 1367 physically bound groups, with a total of 3040 galaxies, plus 14672 galaxy singlets. Most of the galaxies in voids are singlets (59\%), in contrast, most of the NCNV galaxies in the control sample are in groups (60\%). To consider the dynamical stage of the groups we used the parameters harmonic radius ($\rm R_H$), radial velocity dispersion ($\rm \sigma_{v_r}^2$), dimensionless crossing time ($\rm H_0 t_c$), and group virial mass ($\rm  M_{vir}$). We also used the total optical ($r$-band) luminosity, L$_r$, to estimate the mass-to-light ratio ($\rm M/L$) of the groups. We studied the relations of void properties and these parameters with the group richness.} 
   {Galaxy groups can be found in any void in the local Universe, with no dependency of group richness on the density of voids. The densest groups in the studied sample of voids are composed of six galaxies, therefore, voids generally contain small groups, in comparison to denser structures such as filaments, walls, and galaxy clusters. Galaxy groups within voids are typically loose groups, in an early stage of their evolution.}


   \maketitle
%

\section{Introduction}
\label{sec:intro}


Cosmic voids are vast regions in the Universe defined by their unusually low density of galaxies. Surrounded by dense structure (filaments, walls, and clusters), voids represent about 70\% of the volume of the present Universe, being an essential component of the cosmic web \citep{2011IJMPS...1...41V,2017ApJ...835..161M,2024MNRAS.527.4087J}. The formation of voids began with tiny density variations in the early Universe after the Big Bang. Over time, gravity pulled matter from these under-dense regions into the denser areas, leading to the formation of the cosmic web and the expansion of the voids \citep{1996Natur.380..603B}. For this reason, voids are crucial for understanding the overall structure and expansion of the Universe \citep{2007MNRAS.381...41H,2009MNRAS.395.1915T,2013MNRAS.434.1435C, 2017MNRAS.470...85L,2024MNRAS.527.2663L}. The pristine low-density environment of voids provides an ideal place to examine the influence of environment on the formation and evolution of galaxies. 

Despite being the least dense regions of the cosmic web, galaxies do exist within the cosmic voids, although they are much sparser than in denser regions like filaments and clusters \citep{2009MNRAS.400.1105P,2014MNRAS.440L.106A,2020MNRAS.493..899H}. Galaxies residing in voids, the so-called void galaxies, serve as pristine examples for studying galactic evolution, providing insights into how galaxies form and grow with minimal environmental influence \citep{1999AJ....118.2561G,2002A&A...389..405P,2003MNRAS.340..160B,2004ApJ...617...50R}. 
Void galaxies are often less massive, bluer, presenting late-type morphologies, and actively forming stars, likely from pristine intergalactic gas due to the lack of surrounding galaxies that may potentially transform their morphology or halt their star formation, showing low gas-phase metallicities \citep{2000AJ....119...32G,2004ApJ...617...50R,2005MNRAS.356.1155C,2010MNRAS.409..936P,2012MNRAS.426.3041H,2015ApJ...810..165L,2021ApJ...906...97F,2023MNRAS.524.5768P,2023A&A...680A.111D,2023MNRAS.521..916R,2024A&A...692A.258A}. Although there is a small fraction of early-type galaxies in voids, these are systematically smaller (10-20\%) than in denser environments \citep{2025A&A...695A..84P}. Void galaxies might also have had different star formation histories (SFHs): with more continuous and slower star formation rates than in denser environments \citep{2023Natur.619..269D}. Recent integral field spectroscopy (IFS) observations from the Calar Alto Void Integral-field Treasury surveY (CAVITY) project \citep{2024A&A...689A.213P} have revealed detailed spatially resolved stellar population properties of void galaxies \citep{2024A&A...687A..98C}, showing a slightly higher half-light radius, lower stellar mass surface density, and younger ages (across all morphological types) than galaxies in filaments and walls. Moreover, both observations and simulations, have found trends of galaxy properties as a function of the distance to the void centre, hereafter void-centric distance. For instance, an increase of star formation activity and HI gas density has been reported in the inner region of the voids in comparison to larger void-centric distances, which points towards a large-scale modulation of star formation, and higher stellar mass fraction with increasing void-centric distance \citep{2008MNRAS.390L...9C,2014MNRAS.445.4045R,2022MNRAS.517..712R}. Void galaxies may also be connected by less massive void filaments or tendrils \citep{2014MNRAS.440L.106A} and therefore galaxy properties might also be related to their local environment within voids \citep{2000AJ....119...32G,2012A&A...545A.104L,2023A&A...671A.160G,2025arXiv250615345G}. 


Numerical simulations have shown that cosmic voids present an intricate internal network of substructures, making them a more complex environment than expected \citep{2009MNRAS.400.1105P,2010MNRAS.404L..89A,2013MNRAS.435..222R,2014MNRAS.441.2923C,2024MNRAS.527.4087J}. The internal matter distribution within the void (the void environment) might have an impact on the properties and evolution of void galaxies. However, not much work has been done on the observational identification and characterisation of void substructures to understand the impact of the void environment on galaxy evolution.  
\citet{2014MNRAS.440L.106A} quantified the filamentarity (i.e. linearity of structure) within a sample of voids identified in the Galaxy and Mass Assembly (GAMA) survey \citep{2009A&G....50e..12D,2011MNRAS.413..971D} through the GAMA Large Scale Structure Catalogue \citep{2014MNRAS.438..177A}. They identified tendrils as coherent structures composed of up to five or six void galaxies, spanning about 10\,h$^{-1}$\,Mpc, and found them to be rooted in filaments, either connected to other filaments or terminating in the voids. However, it is unknown if the galaxies are physically bound in groups within these tendrils. \citet{2025A&A...700A.196C} compared galaxy pairs in voids with those in denser structure, finding increased star formation activity in void paired galaxies, which are also bluer. In addition, \citet{2022A&A...665A..44A} characterised the halo occupation distribution of groups in cosmic voids, reporting that for the low-mass groups the youngest galaxies are only present inside voids, generally as central galaxies, where haloes populating voids have had a different formation history, inducing significant changes on the halo occupation distribution \citep{2020A&A...638A..60A}.
\citet{2024A&A...691A.341T} went one step further and classified galaxies as singlets and physically bound group members in voids, using the \citet{2012MNRAS.421..926P} catalogue of cosmic voids and void galaxies from the Sloan Digital Sky Survey Data Release 7 \citep[SDSS-DR7;][]{2009ApJS..182..543A}, to investigate how the local environment influences the SFH of galaxies in comparison to galaxies in walls and filaments, and clusters. They found that the large-scale environment is related to a delay in mass assembly of up to $\sim$2\,Gyr, while the local environment is related to a shorter delay ($<$1\,Gyr). However, no identification of individual groups is provided to study their properties as a function of void peculiarities, or the properties of the void galaxies composing the groups. The degree to which the location of void galaxies within cosmic voids affects their evolution, considering the complexity of the local environment within the void, is an open question.


Finding galaxy groups in voids is not an easy task. Results from simulations have shown that, in contrast to denser environments, cosmic voids are more devoid of galaxies than they are of mass \citep{2024ApJ...962...58C}. This means that the density profiles and tools generally used to identify galaxy groups and clusters in the large-scale structure \citep[e.g. halo-based group finder algorithms;][]{2007ApJ...671..153Y} might not apply within voids or could provide biased results. Catalogues of groups and clusters based on friend-of-friends (FoF) algorithms, are usually limited to groups with a minimum richness $\geq$3-4 galaxies, or even $\geq$8-10 galaxies \citep{2002MNRAS.335..216M,2007A&A...474..783D,2008A&A...479..927T,2009ApJS..183..197W}, which under-represents the void environment. In other cases, the analysis of the groups is limited to denser groups, due to the uncertainties at lower densities \citep{2017A&A...602A.100T}. A mixed methodology, i.e. finding physically bound galaxies within their local environment, would provide more accurate results. For instance, \citet{2023MNRAS.520.6367T} applied a Hickson-like criteria to find compact groups (CGs) in different large-scale environments, finding almost no CGs residing in the inner regions of cosmic voids. In the present work, we aim to identify and characterise galaxy groups within voids in the \citet{2012MNRAS.421..926P} catalogue of voids, taking into account the peculiarities of these vast and empty structures in the local Universe ($z$\,<\,0.08). To do so, we applied a FoF-like group finder algorithm, ensuring a certain degree of gravitational bounding among group members. For this, we followed the criteria to select physically bound neighbours in isolated systems in \citet{2015A&A...578A.110A}, instead of using a fixed aperture or a n$\rm ^{th}$ neighbour approach. To characterise these groups, assuming they are virialised, we estimate their dynamical properties and group mass-to-light ratios to investigate how light is traced by the underlying distribution of matter in voids.

This work is organised as follows. In Sect.~\ref{sec:data} we describe the sample of voids and void galaxies used in this work. In Sect.~\ref{sec:method} we describe the procedure to identify galaxy groups within voids and derive the properties to characterise these systems. In Sect.~\ref{sec:res} we show the main results, and discuss them in Sect.~\ref{sec:dis}. Finally, in Sect.~\ref{sec:con} we present the main conclusions of this work. Throughout the paper, we use a cosmology with $\Omega_{\Lambda 0} = 0.7$, $\Omega_{\rm m 0} = 0.3$, and $\rm H_0 = 70$\,km\,s$^{-1}$\,Mpc$^{-1}$.

\section{Void galaxy sample} \label{sec:data}

The sample of void galaxies that we use in this study is based on a well-defined sample of void galaxies, from which the parent sample of the CAVITY\footnote{\url{https://cavity.caha.es/}} project \citep{2024A&A...689A.213P} was built. CAVITY is a legacy project aimed at characterising the population of galaxies inhabiting voids using IFS data \citep{2024A&A...691A.161G,2025hsa..conf...93P}. The CAVITY parent sample is based on the catalogue of cosmic voids and void galaxies compiled by \citet{2012MNRAS.421..926P} using photometric and spectroscopic data from the SDSS-DR7 \citep{2009ApJS..182..543A}. 
\citet{2012MNRAS.421..926P} used a galaxy-based void-finding algorithm that uses redshift data to objectively find voids using a nearest-neighbour algorithm on a volume-limited galaxy catalogue. This VoidFinder algorithm is described by \citet{2002ApJ...566..641H}, based on the original method by \citet{1997MNRAS.287..790E}. The algorithm uses the three-dimensional distribution of galaxies in the SDSS-DR7 to initially classify galaxies as wall or void galaxies, understanding as void galaxies the galaxies that reside in a void region, whereas "wall" galaxies reside in walls, filaments, and clusters. To do so, \citet{2012MNRAS.421..926P} estimated the projected distance to the third nearest neighbour ($\rm d_{3rd}$) of each galaxy. Galaxies are classified as wall galaxies if $\rm d_{3rd}$\,$<$\,7\,Mpc. If distance is larger, they are left as potential void galaxies. In a first step, only the distribution of wall galaxies is used to identify the voids, using the VoidFinder algorithm to grow empty cells as spheres (merging the spheres when their overlap exceeds 50\%). A void is defined when their growing sphere is bound by four wall galaxies, with their centre not confined to the initial cell, with a cut-off of 10\,$h^{-1}$\,Mpc for the minimum radius of a void region. 
The sizes of the voids ($\rm R_{void}$) are measured using their effective radius ($\rm r_{eff}$), this is, the radius of voids (if spherical) or the radius of the sphere that has the same volume as the non-spherical void (if not spherical). The voids in the \citet{2012MNRAS.421..926P} catalogue have a median effective radius of $17.83\,h^{-1}$\,Mpc, with values ranging from $10\,h^{-1}$\,Mpc (by definition) to $30\,h^{-1}$\,Mpc.
In a second step, void galaxies are selected from the list of candidates as those falling within the limits of the identified voids. \citet{2012MNRAS.421..926P} identified 1054 statistically significant voids covering 62\% of the volume of the Universe up to $z$\,$<$\,0.107. These voids are populated by 8046 galaxies brighter than an absolute magnitude limit of M$_{r}\,<\,-20.09$\,mag (the volume-limited galaxy sample) and 79947 void galaxies with $\rm m_{r}\,\leq\,$17.7\,mag and $z$\,<\,0.088 (the magnitude-limited galaxy sample). 

The design of the CAVITY sample secures a comprehensive investigation of void galaxies spanning various void sizes and dynamical stages. The target selection is based on the magnitude-limited sample in order to have a statistically robust sample size, while restricting to voids in the catalogue where the full volume (projected) of the void is included in the SDSS footprint. The CAVITY parent sample is limited to voids in the redshift range 0.005 and 0.050, to ensure an adequate spatial coverage across galaxy disks. For the purpose of the present study, we follow the same selection criteria but increasing the redshift range up to $z$\,$<$\,0.08. This leaves us with 24931 void galaxies in 170 voids.  

\begin{table}
    \centering
    \caption{\label{tab:grichness}Group richness and number of singlets and groups.}
    \begin{tabular}{ccccc}
    \hline\hline
    & $\rm N_{gal}$ & $\rm N_{groups}^{full}$  &  $\rm N_{groups}^{lim}$ & $\rm N_{groups}^{vol}$ \\ 
    \hline
    Voids & 1 & 14672 & 1013\tablefootmark{a} & 7674 \\
    & 2 & 1137  & 292                   & 128  \\
    & 3 & 175   & 75                    & 10   \\
    & 4 & 36    & 18                    & 5    \\
    & 5 & 17    & 9                     & --   \\
    & 6 & 2     & 1                     & --   \\
    \hline
    NCNV & 1 & 44503 & 5082 & 27795 \\
    & 2 & 20748 & 7062 & 18048 \\
    & 3 & 7737  & 4034 & 8042  \\
    & 4 & 2992  & 1996 & 3491  \\
    & 5 & 1459  & 1161 & 1395  \\
    & 6 & 626   & 542  & 545   \\
    \hline
     \end{tabular}
    \tablefoot{Number of groups in voids and in the control sample of NCNV groups} when considering the full sample ($\rm N_{groups}^{full}$) or the limited sample ($\rm N_{groups}^{lim}$) as a function of the group richness ($\rm N_{gal}$). For reference, the richness for groups in a volume-limited sample ($\rm N_{groups}^{vol}$) with $\rm M_{r}\,\leq\,$-20.1\,mag, up to $z$=0.08, is also shown. \tablefoottext{a}{Among singlets in the full sample in voids, there are 1013 potentially isolated galaxies (singlets in the limited sample).}
\end{table}

\begin{figure*}
\begin{center} 
\includegraphics[width=0.4\textwidth]{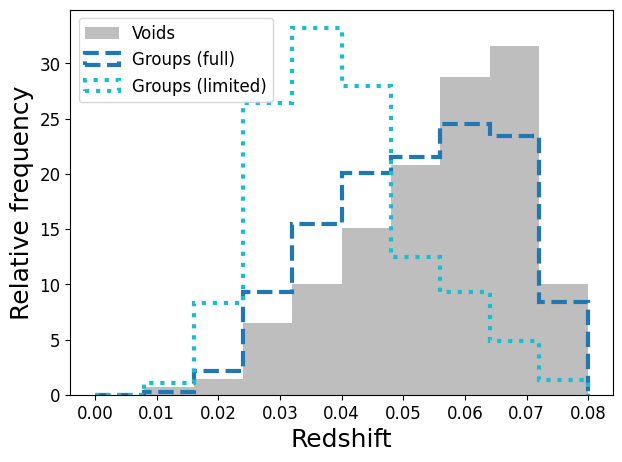}
\includegraphics[width=0.49\textwidth]{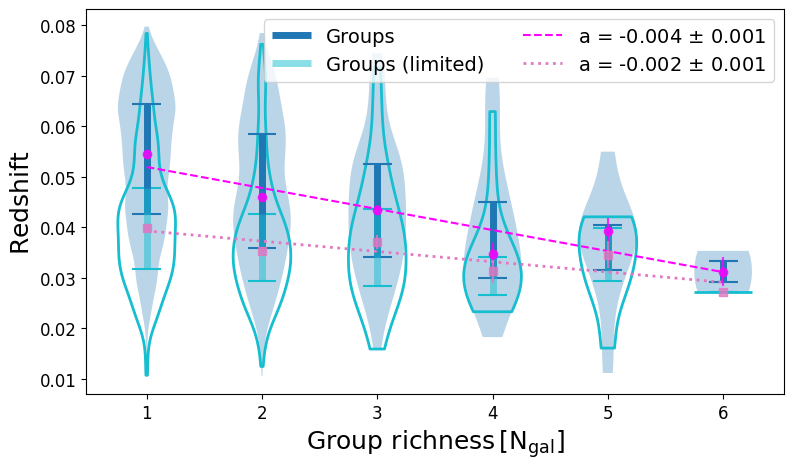}
\caption[]{Left panel: Redshift distribution of the groups, as blue dashed and cyan dotted open histograms, for the full (14672 singlets and 1367 groups) and limited (1013 singlets and 395 groups) sample, respectively, in comparison to the distribution of the redshift of the voids (gray filled histogram). Right panel: redshift of the group as a function of the group richness. The redshift distribution for groups with same group richness is represented by violin plots (blue filled distributions considering the full sample, and cyan empty distributions for the limited sample). The inner box in each violin plot represents the interquartile range of the median and its 95\% of confidence intervals. Magenta circles and pink squares correspond to the median values in richness bins (considering the full sample or the limited sample, respectively) with their corresponding uncertainties, while dashed magenta line and dotted pink line correspond to the linear least-square fit to the binned data, respectively. The slopes of the linear fits, with their corresponding uncertainties, are indicted in the legend.} \label{fig:zgroups} 
 \end{center}
\end{figure*}


\section{Galaxy groups and properties} \label{sec:method}

\subsection{Identification of physically bound groups} \label{sec:pipeline}

Each void is evaluated independently, considering only the galaxies that reside in it, in a two iteration process. First, we look for neighbour galaxies with line-of-sight velocity difference $\rm \Delta v\,\leq\,\pm~500\,km\,s^{-1}$ within 1\,Mpc in projected distance \citep{2015A&A...578A.110A}. Galaxies with no neighbours within that volume are automatically classified as singlets (galaxies that are not physically bound to any other detected galaxy). A total of 14672 galaxies are classified as singlets. For the galaxies with neighbours in that volume, we follow a similar methodology as in \citet{2024A&A...691A.341T} to find grouped galaxies in voids, which used the criteria by \citet{2015A&A...578A.110A} to find physically bound neighbours in low density environments (i.e. isolated galaxies and isolated pairs). Physically bound neighbours are galaxies separated by a projected distance d\,$\leq$\,450\,kpc with $\rm \Delta v\,\leq\,\pm~160\,km\,s^{-1}$. The first cut identifies candidate neighbours, while the second ensures physical bounding. In this first iteration, a sample of 2373 group candidates, with more than one neighbour, is identified. The central galaxy of each group candidate is identified as the most massive galaxy (see Sect.~\ref{sec:ML} for the details of the estimation of the stellar mass). We use this definition to identify unique groups in a second iteration as follows. 

Using this methodology, there might be different groups, even with different number of members, with the same central galaxy. In this case we consider only the group with the largest number of galaxies. Also, some member galaxies in a group can be central galaxies of another nearby group. This means that the two groups are physically connected. When this happens, we merge the two groups and identify the central galaxy as the most massive of both groups. When a group member is found in two different groups but is not identified as the central galaxy in any of them, we consider the two groups separately, even if they share neighbours, since these galaxies would not physically connect the two groups. After this second iteration, a sample of 1367 physically bound groups, composed of 3040 galaxies, are identified in the sample of 170 voids considered in this work, where the groups with the largest number of members are composed of six galaxies. 
The typical number of members in the groups, hereafter group richness ($\rm N_{gal}$), is $\rm N_{gal}$\,=\,2.

Note that, even if singlet galaxies do not have any neighbour within $\rm \Delta v\,\leq\,\pm~500\,km\,s^{-1}$ at 1\,Mpc projected distance, this does not necessarily mean that singlet galaxies are isolated. Isolated galaxies have not been appreciably affected by their closest neighbours during a past crossing time t$_{cc}\,\approx\,$3\,Gyr \citep{2005A&A...436..443V,2007A&A...472..121V,2007A&A...470..505V}. Considering that the redshift completeness of the SDSS main galaxy sample is flux limited to an apparent magnitude $\rm m_{r}\,=\,$17.77\,mag, the faintest singlets at higher redshift might look as isolated simply because possible physically bound neighbours are even fainter, therefore these galaxies do not have an observed spectrum and thus, are not considered in our analysis. Following the criteria in \citet{2015A&A...578A.110A}, we can identify a sample of potential isolated galaxies from the sample of singlets when $\rm m_{r}\,\leq\,$15.7\,mag, which allows an homogeneous isolation definition without neighbours within at least 2 magnitudes fainter than the singlets. The subsample of potential isolated galaxies is composed of 1013 void galaxies, $\sim$7\% of the singlets. According to the criteria in \citet{2015A&A...578A.110A}, these galaxies would have been "passively evolving" over the last $\sim$5\,Gyrs (since z$\sim$1). Note that ultra-deep optical imaging of these galaxies may reveal faint external features that might indicate a probable history of interaction \citep{2023A&A...677A.117S}. 

Following the same motivation, i.e. having groups with neighbours up to 2 magnitude fainter than central galaxies, and to take into account any potential bias or dependency with the redshift in the analysis of the results, we defined a subsample of groups where the central galaxy has $\rm m_{r}\,\leq\,$15.7\,mag. Hereafter we refer to this subsample as the limited sample. Previous works by \citet{2007ApJ...671..153Y} or \citet{2017A&A...602A.100T}, for instance, have also defined a volume-limited subsample within a redshift range to test any potential redshift bias. For reference, we have also compiled a subsample of groups with galaxies with absolute magnitude $\rm M_{r}\,\leq\,-20.1$\,mag, up to $z$\,=\,0.08. However, throughout our analysis, we use the limited sample, instead of the absolute magnitude volume-limited sample for all group members, to consider any redshift dependence in comparison to the full sample, because of two main reasons. First, an absolute magnitude volume-limited sample under-represents low redshift galaxies ($z$\,<\,0.02), and second, it also under-represents the characteristic population of galaxies in voids, which are often less massive and with lower surface brightness than galaxies in denser environments \citep{2011ApJ...728...74G,2021ApJ...906...97F,2024A&A...687A..98C}, therefore we would be missing the main galaxy populations in voids, including dwarf galaxies \citep{2025A&A...698A.260B,2025A&A...693L..16B}. In fact, the maximum richness when using a volume-limited sample (with absolute magnitude $\rm M_{r}\,\leq\,$-20.1\,mag, up to $z$=0.08) is $\rm N_{gal}$=4 for a total of five groups, reducing the study of the dynamical properties of the groups to three richness bins, limited to groups with $\rm N_{gal}$=2, 3, and 4, which have low statistics. 
Table~\ref{tab:grichness} shows the number of groups for each richness for the full, limited, and volume-limited samples. The left panel of Fig.~\ref{fig:zgroups} shows the  redshift distribution for the full and limited samples in comparison to the redshift of the voids (grey filled histogram), measured as the median the redshift of the galaxies that reside in each void. The number of voids increases at higher redshift because the comoving volume spanned by the survey is larger. While the full sample follows the distribution of the voids, the distribution of the limited sample peaks at $z$\,$\sim$\,0.035, following the redshift distributions for isolated galaxies, pairs, and triplets in \citet{2015A&A...578A.110A} (as shown in their Fig.~2.). 
The right panel of Fig.~\ref{fig:zgroups} shows that group richness decreases with the redshift. This trend is steeper for the full sample than for the limited sample, showing that most of the singlets in the full sample are found at higher redshift, being affected by redshift bias. However, the differences are minimal (within the uncertainties) at higher group richness. 

\begin{figure}
\begin{center} 
\includegraphics[width=0.9\columnwidth]{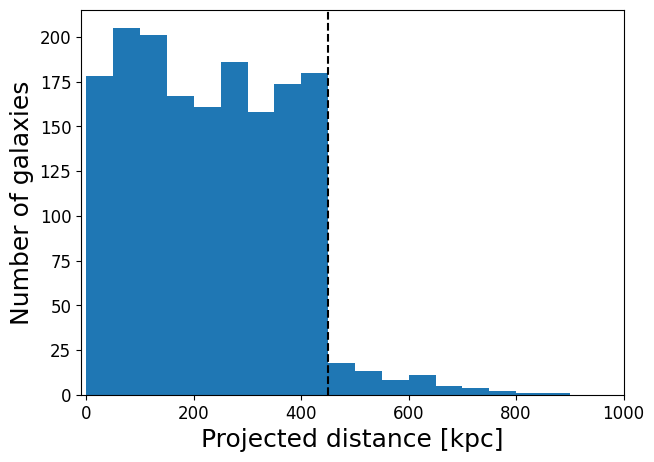}\\
\includegraphics[width=0.9\columnwidth]{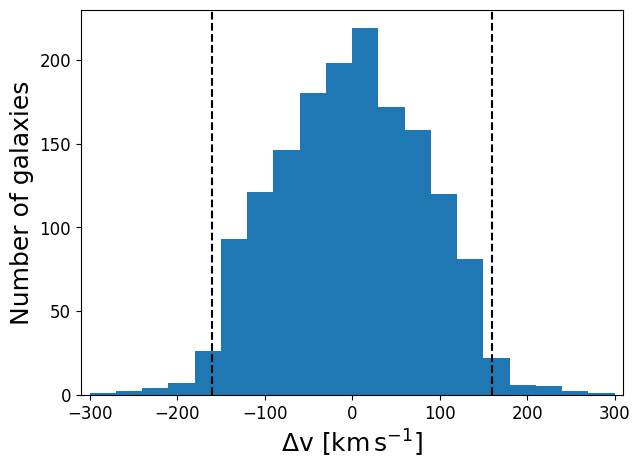}\\
\caption[]{Distribution of the projected distance (upper panel), in kpc, and line-of sight velocity difference (lower panel), in $\rm km\,s^{-1}$, of group members with respect to the central galaxy of the group (full sample), for groups with $\rm N_{gal}\,\geq\,2$. The vertical black dashed lines indicate the limit values used to define the physically bound groups in the first iteration, following \citet{2015A&A...578A.110A}.} \label{fig:distvel} 
 \end{center}
\end{figure}

Figure~\ref{fig:distvel} shows the distribution of the projected distance and line-of-sight velocity difference for group galaxies with respect to their central galaxy, for groups with more than one physically bound members. As shown in the figure, most of the group members are located within the limit values used to define physically bound neighbours. As an effect of the second iteration, a small fraction of group members can be found at projected distances up to 890\,kpc, and with a $\rm \Delta v$ not larger than $\rm \pm~300\,km\,s^{-1}$. Note that these two conditions not necessarily occur simultaneously for the same member galaxy, and do not exclude these groups from the analysis. While the adopted linking criteria ($\Delta v\,\leq\,\pm~160\,km\,s^{-1}$, d\,$\leq$\,450\,kpc) are conservative for low-density voids and follow established methods for physically bound pairs \citep{2015A&A...578A.110A}, we note that Hubble flow contributions ($\sim$\,35\,km\,s$^{-1}$ at 500\,kpc for $\rm H_0 = 70$\,km\,s$^{-1}$\,Mpc$^{-1}$) and extended distances post-merging ($\sim$\,890\,kpc in rare cases) may introduce minor contamination by non-bound projections.

\subsection{Control sample of groups}
\label{sec:CS}

To compare with the properties of the population of galaxy groups in voids, we have built a control sample of galaxy groups outside voids. To do so, we have applied the same algorithm to identify galaxy groups in voids, but to identify groups in a sample of galaxies not belonging to clusters or voids, hereafter NCNV (not in voids nor clusters), following \citet{2026arXiv260500982T}. To select NCNV galaxies, we used the catalogue of group galaxies classified by \citet{2017A&A...602A.100T}, composed of 584449 galaxies up to a redshift $z$\,=\,0.2, based on data from the SDSS-DR12 \citep{2011AJ....142...72E,2015ApJS..219...12A}. First of all, we restricted the sample to galaxies in the same redshift range as the sample of 170 voids used in this study ($z$\,<\,0.08, 220614 galaxies). From this, we removed cluster galaxies \citep[i.e., galaxies being part of groups of 30 galaxies or more,][]{1989ApJS...70....1A}, and void galaxies \citep[i.e. galaxies in common with the catalogue of void galaxies,][]{2012MNRAS.421..926P}, leaving us with 156700 NCNV galaxies. 

After applying the group finder algorithm, we identified 44503 singlets ($\sim$\,28\% of NCNV galaxies) and 11121 groups composed of 93470 galaxies (60\% of NCNV galaxies). To use this sample as a control sample, we only consider galaxy groups with similar richness (up to $\rm N_{gal}$\,=\,6), leaving us with 33562 groups composed of 87726 galaxies. Similarly as for groups in voids, we identified a limited sample and a volume-limited sample. The number of groups for each richness in each sample is also shown in Table~\ref{tab:grichness}.

\subsection{Total stellar mass and optical luminosity}
\label{sec:ML}

For each group, we estimated their total stellar mass and optical luminosity in the $r$-band as the sum of the stellar mass and luminosity of their member galaxies (including singlets). Stellar masses were estimated by fitting the spectral energy distribution on the five SDSS bands using the k-correct code \citep{2007AJ....133..734B}. Luminosities were calculated using magnitudes corrected for extinction in the $r$-band. Galactic reddening corrections in magnitudes at the position of each object are computed following \citet{1998ApJ...500..525S}. 
To avoid selection effects due
to the slightly different redshift of the group members, and to be consistent with the methodology to identify galaxy groups, we consider the redshift of the central galaxy as the redshift of the group. 

\begin{figure}
\begin{center} 
\includegraphics[width=0.9\columnwidth]{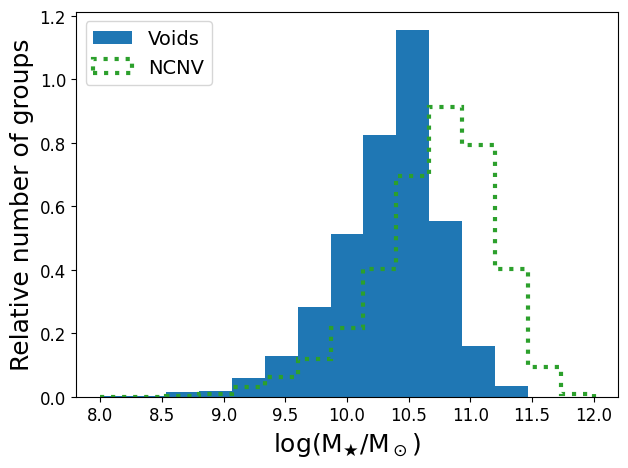}\\
\includegraphics[width=0.9\columnwidth]{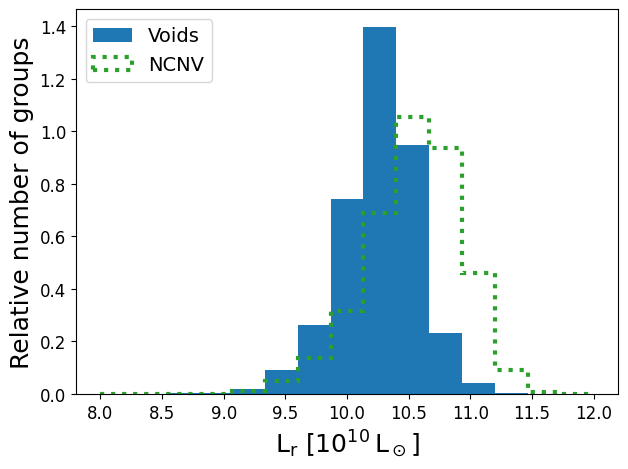}\\
\caption[]{Distribution of the total stellar mass (upper panel), and total optical luminosity (lower panel) for the galaxy groups and singlets in the full sample, in solar units. The distributions for the control sample of NCNV groups are presented as green dotted open histograms.} \label{fig:masslum} 
 \end{center}
\end{figure}

The distributions of the computed total stellar mass and total luminosity of each group are presented in Fig.~\ref{fig:masslum}. 
The median value of the total stellar mass of the groups is $\rm \log(M_\star/M_\odot)$\,=\,10.4$_{10.1}^{10.6}$, while for the total optical luminosity is $\rm \log(L_r/L_\odot)$\,=\,10.3$_{10.1}^{10.5}$; where the upper and lower indices correspond to the 25$^{th}$ and 75$^{th}$ percentiles of the corresponding distribution, respectively. These values increase to $\rm \log(M_\star/M_\odot)$\,=\,10.8$_{10.5}^{11.0}$ and $\rm \log(L_r/L_\odot)$\,=\,10.6$_{10.3}^{10.8}$ in the control sample of NCNV groups.

\subsection{Dynamical parameters}
\label{sec:dynpar}

To characterise the dynamical stage of the galaxy groups, we calculate their projected harmonic radius $\rm R_H$, radial velocity dispersion $\rm \sigma_{v_r}^2$, dimensionless crossing time $\rm H_0 t_c$, and group virial mass $\rm  M_{vir}$, defined as follows:

\begin{equation} \label{eq:eq1}
  \rm  R_H = \left( \frac{1}{N_{gal}}\sum R_{ij}^{-1}\right)^{-1}\,,
\end{equation}

\begin{equation} \label{eq:eq2}
 \rm   \sigma_{v_r}^2 = \frac{1}{N_{gal}-1}\sum \left(v_r - \langle v_r \rangle \right)^2\,,
\end{equation}

\begin{equation} \label{eq:eq3}
  \rm  H_0 t_c = \frac{H_0\,\pi \,R_H}{\sqrt{3}\,\sigma_{v_r}}\,,
\end{equation}
and

\begin{equation} \label{eq:eq4}
  \rm  M_{vir} = \frac{3\pi\, N_{gal}\, R_H\, \sigma^2_{v_r}}{\left(N_{gal}-1\right)G}\,,
\end{equation}
where $\rm R_{ij}$ are the projected galaxy-galaxy separations (in Mpc), $\rm N_{gal}$ is the group richness, $\rm v_r \propto c z$, and $\rm \langle v_r \rangle$ the mean value of the radial velocity considering all member galaxies of the group.
These parameters, as defined in \citet{1982ApJ...255..382H} and \citet{1992ApJ...399..353H}, quantify the dynamic configuration of groups and clusters. They provide a robust estimation of the evolution stage on systems, even in groups with a low number of members \citep{2016RAA....16...72F,2023A&A...670A..63V}.
The $\rm R_H$ parameter represents a measurement of the effective radius of the galaxy group. The $\rm \sigma_{v_r}^2$ parameter corresponds to the dispersion of the radial velocities of the galaxies in the groups. We use $\rm H_0 t_c$ as a measure of the time that takes a single galaxy to go across the system. The $\rm  M_{vir}$ parameter is the mass estimation of a gravitationally bound system. It is derived from the effective radius of the group and its velocity dispersion, with a factor to account for the group richness. This parameter represents an approximation of the dark matter halo and baryonic mass.

Note that, by definition, these parameters (except $\rm R_H$) can only be computer for groups (i.e. singlets are excluded). 
The de-projection factor $\sqrt{3}$ in Equations~\ref{eq:eq3} and \ref{eq:eq4} is introduced as an estimator of the "true" three-dimensional velocity dispersion of a group. This factor comes from the assumption that groups have isotropic velocity distributions. This assumption may introduce statistical uncertainties considering that the small number of galaxies might dominate the estimate of the velocity dispersion \citep{1993AJ....105.2035D}. We take into account this fact in the analysis of the dynamical parameters.

The group virial mass is also used to compute the total mass-to-light ratio of the groups (M/L), as described in Sect.~\ref{sec:res_ML}. The M/L is a key parameter for inferring the amount of dark matter in groups. However, given the uncertainties, it is impossible to provide a precise value for the M/L of small groups. 
Therefore, the aim of this work is limited to the study of the dependency of the estimated M/L values with the group richness, in comparison with the observed trends in groups located in denser environments in the literature.

\section{Results} \label{sec:res}

\subsection{Galaxy groups in voids} 

Following the procedure described in Sect.~\ref{sec:pipeline}, we have identified a sample of group galaxies within cosmic voids in the local Universe ($z$\,<\,0.08). The list of galaxies in groups (in the full sample) is provided in Table~\ref{tab:galaxies}, while the catalogue of groups is provided in Table~\ref{tab:groups}. The catalogue of groups is composed of 1367 physically bound groups (for a total of 3040 galaxies), plus 14672 galaxy singlets (galaxies that are not physically bound to any other detected galaxy). These groups are distributed in 170 voids from the catalogue of voids in \citet{2012MNRAS.421..926P}, from which the CAVITY parent sample was built. In contrast to void galaxies, where most of the galaxies are singlets (59\%), most of the galaxies in the NVNC control sample are in groups (60\%). As shown in Table~\ref{tab:grichness}, the densest groups in voids are composed of six galaxies. The typical richness of voids is $\rm N_{gal}\,=\,2$ (83\% of the groups in the full sample and 74\% in the limited sample). Therefore, voids generally contain small groups, in comparison to denser structures such as filaments and walls, with approximately $\sim$64\% of pair galaxies \citep{2020A&A...639A..71K}, and galaxy clusters \citep{2017A&A...602A.100T}. 
In particular, 62\% of the groups in the control sample of NVNC groups are composed of two physically bound galaxies, in agreement with the literature.
To investigate how our algorithm performs in comparison to other group finder algorithms, we have done a crossmatch with the galaxies in the catalogue of groups in \citet{2017A&A...602A.100T} (FoF algorithm) and in the catalogue of \citet{2007ApJ...671..153Y} (halo-based algorithm) that are also found in the catalogue of void galaxies \citep{2012MNRAS.421..926P}. In general, there is a good agreement: 83\% of singlets in our catalogue are also singlets with a FoF algorithm (81\% when considering the limited sample), while 94\% of the group galaxies are still group galaxies (91\% when considering the limited sample). When comparing with a halo-based algorithm, 73\% and 67\% are still singlets and group galaxies, respectively (91\% and 78\%, respectively, when considering the limited sample). Note that \citet{2007ApJ...671..153Y} applied a strict magnitude limit cut-off for group members, and therefore there is a better agreement with the limited sample than for the full sample.

\begin{figure}
\begin{center} 
\includegraphics[width=\columnwidth]{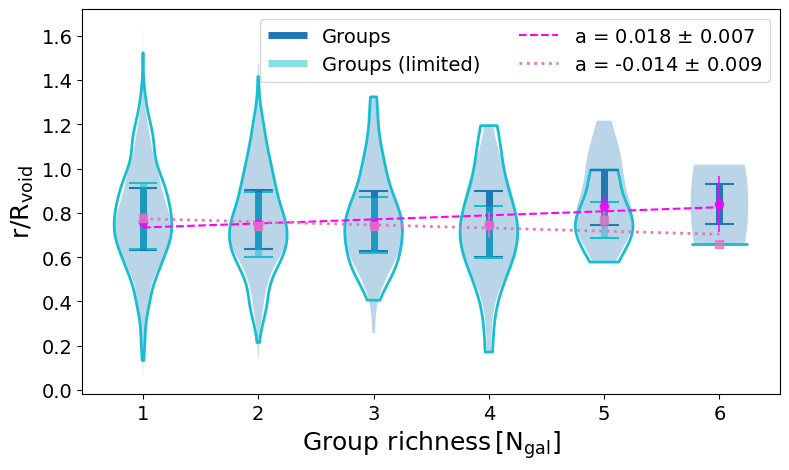}
\caption[]{Void-centric distance as a function of group richness. A void-centric distance value of the unit represents the maximum size of the voids if they were spherical. The distribution of void-centric distance for groups with the same group richness is represented by violin plots (blue filled distributions considering the full sample, and cyan empty distributions for the limited sample). The inner box in each violin plot represents the interquartile range of the median and its 95\% of confidence intervals. Magenta circles and pink squares correspond to the median values in richness bins (considering the full sample or the limited sample, respectively) with their corresponding uncertainties, while dashed magenta line and dotted pink line correspond to the linear least-square fit to the binned data, respectively. The slopes of the linear fits, with their corresponding uncertainties, are indicted in the legend.} \label{fig:fracR} 
 \end{center}
\end{figure}

In Fig.~\ref{fig:fracR} we present the void-centric distance, r/$\rm R_{void}$, as a function of the group richness. The values of r/$\rm R_{void}$ were computed after normalising the position of the central galaxy in each group by the void size, $\rm R_{void}$. A value of r/$\rm R_{void}$=1 represents the maximum size of the voids if they were spherical. As it can be appreciated from the figure, galaxy groups are evenly distributed within voids, however the densest groups in voids avoid the innermost regions (r/$\rm R_{void}$<0.6). We note that the environment within voids might be an intricate region. Voids are not completely spherical, they are composed of the superposition of various cavities, and therefore, even considering the restricted selection criteria of void galaxies in \citet{2012MNRAS.421..926P}, there might be contamination by wall galaxies inside the void at larger void-centric distances. Also, as noted by \citet{2007ApJ...671..153Y} and \citet{2017A&A...602A.100T}, galaxy groups can merge with other groups. Hence, dense galaxy groups within voids, and therefore more massive (both in stellar and virial mass), at larger void-centric distances, might be attracted by groups which belong to the walls surrounding voids. To quantify these effects, we used the classification of the environment in \citet{2024A&A...691A.341T}, who used the catalogue of void galaxies in \citet{2012MNRAS.421..926P} to find grouped galaxies in voids, and the catalogue of groups and clusters compiled by \citet{2017A&A...602A.100T} to reduce the contamination by galaxies in walls and filaments (hereafter wall). We found that 37 singlets (0.3\% of the sample) and 39 groups (less than 3\% of the groups) might be associated with wall galaxies, with typical void-centric distance r/$\rm R_{void}$\,=\,0.9 and similar mass (both stellar and virial) than the rest of the groups. This might be related with the observed result that dense groups avoid the inner region of the voids. While voids expand, galaxy groups are pulled towards their surrounding walls. In addition, \citet{2026arXiv260205117A}, based on simulations, has recently reported that the radial gradient of individual bias (i.e. how individual galaxies trace the underlying dark matter field) within voids, generally increasing from negative values at the void centre to higher values at the edges, is robust across most void definitions, including the VoidFinder algorithm used in \citet{2012MNRAS.421..926P}. In contrast, \citet{2026arXiv260329706Z}, based on both observations and simulations, reported that void size distributions and radial density profiles depend strongly on the identification algorithm. The radial number density profiles differences are noticeable at larger void-centric distances (r/$\rm R_{void}$\,$\gtrsim$\,1.0). Therefore, our findings should be consistent, independently of the chosen void catalogue, at least in the inner regions of the voids.

\begin{figure*}
\begin{center} 
\includegraphics[width=0.45\textwidth]{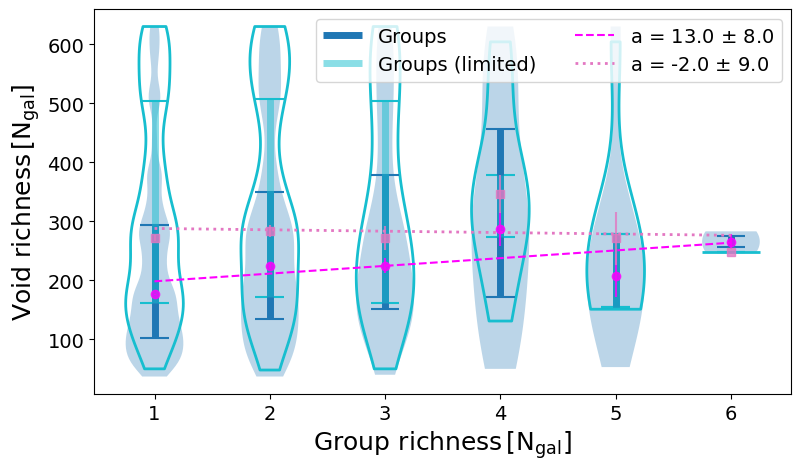}
\includegraphics[width=0.45\textwidth]{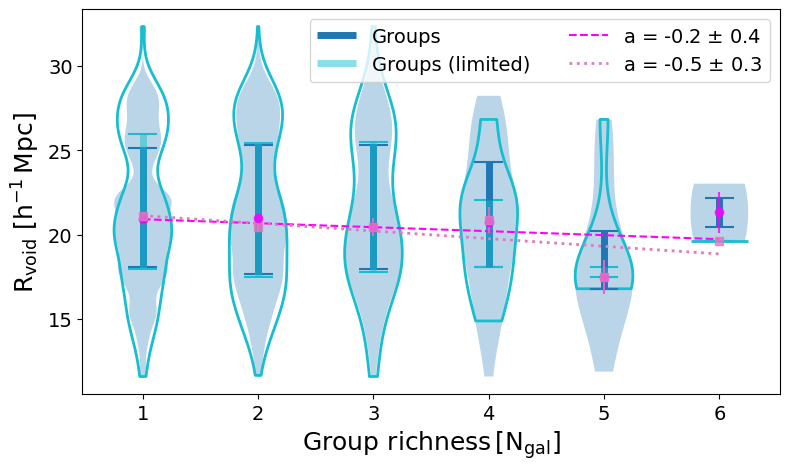}\\
\includegraphics[width=0.45\textwidth]{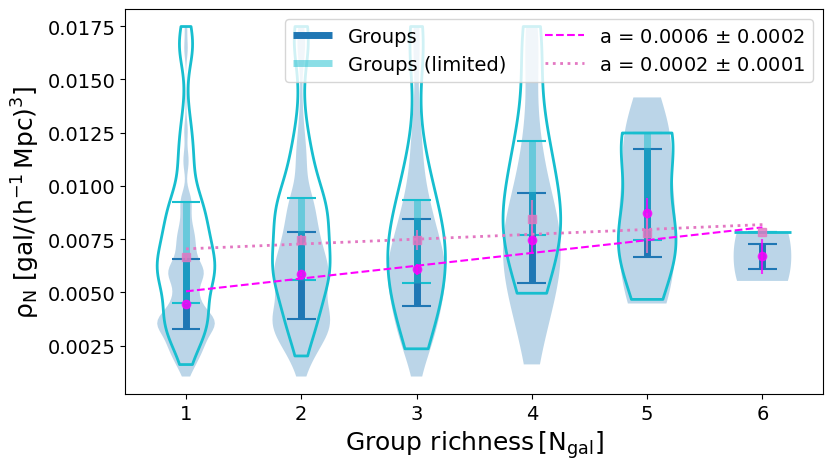}
\includegraphics[width=0.45\textwidth]{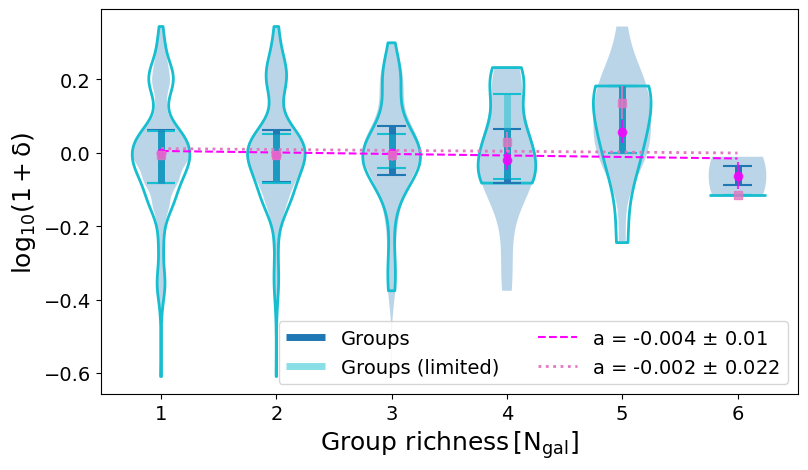}
\caption[]{Properties of the voids as a function of the group richness: number of galaxies in the voids (void richness, top left panel); size of the voids ($\rm R_{void}$, top right panel); the number density of galaxies in each void ($\rm \rho_N$, bottom left panel); and the over-density-corrected number density ($\rm \log_{10}(1 + \delta)$, bottom right panel). The distribution of void properties for groups with same group richness is represented by violin plots (blue filled distributions considering the full sample, and cyan empty distributions for the limited sample). The inner box in each violin plot represents the interquartile range of the median and its 95\% of confidence intervals. Magenta circles and pink squares correspond to the median values in richness bins (considering the full sample or the limited sample, respectively) with their corresponding uncertainties, while dashed magenta line and dotted pink line correspond to the linear least-square fit to the binned data, respectively. The slopes of the linear fits, with their corresponding uncertainties, are indicted in the legend.} \label{fig:voidprop} 
 \end{center}
\end{figure*}

The panels in Fig.~\ref{fig:voidprop} show the number of galaxies in the voids (void richness), void size, and void galaxy number density (including over-density-corrected number density), as a function of the group richness. We observe that group richness is independent of void size (upper right panel), where the median of the distribution of void size at each group richness is similar to the median value for the sample of 170 voids used in this work ($\rm R_{void}$\,$\sim$\,20\,h$^{-1}$\,Mpc). Note that void size does not vary with redshift in the considered redshift range (0.005\,$\leq$\,z\,$\leq$\,0.080). The group richness shows a tendency to increase with void richness (upper left panel), and consequently with galaxy number density of the void (lower left panel), that disappears when considering the limited sample of groups. This is because the void richness is affected by the Malmquist bias that is implicit in the SDSS galaxy sample, which leads to the preferential detection of intrinsically bright galaxies at higher redshift. For this reason we also consider the over-density-corrected galaxy number density of voids ($\rm log_{10}(1 + \delta)$, shown in the lower right panel), as defined in \citet{2024A&A...692A.258A}, in which the over-density for the void-galaxy number density is estimated in each void using the methodology of \citet{2015MNRAS.451..660E}. As a result, we find that the group richness is independent of the density of the voids. This implies that there are no voids composed only of singlets or with no groups.

\begin{table*}
    \centering
    \caption{\label{tab:galaxies}Catalogue of group galaxies.}
\begin{tabular}{cccccccc}
\hline\hline
(1) & (2) & (3) & (4) & (5) & (6) & (7) & (8) \\
groupID & voidID & cavityID & RA & Dec & z & $\rm m_r$ & $\rm M_r$ \\
        &        &          & (deg) & (deg) &  & (mag) & (mag) \\
\hline\hline
1 & 44 & 1754 & 133.959488 & 0.831134 & 0.041743 & 15.44 & $-$21.15 \\
1 & 44 & 2884 & 133.859375 & 0.858172 & 0.041981 & 17.19 & $-$19.42 \\
2 & 649 & 2003 & 150.803421 & 1.148881 & 0.067632 & 16.46 & $-$21.09 \\
2 & 649 & 2004 & 150.816711 & 1.144060 & 0.067632 & 16.46 & $-$21.09 \\
3 & 649 & 2015 & 154.869186 & 1.062658 & 0.065402 & 16.68 & $-$20.87 \\
3 & 649 & 69946 & 154.849915 & 1.052281 & 0.065557 & 17.66 & $-$19.90 \\
4 & 349 & 2859 & 143.087585 & 0.506555 & 0.071858 & 17.56 & $-$20.22 \\
4 & 349 & 69991 & 143.024017 & 0.453879 & 0.072006 & 17.72 & $-$20.07 \\
5 & 38 & 3137 & 127.461372 & 48.374413 & 0.054634 & 16.15 & $-$20.96 \\
5 & 38 & 70026 & 127.411751 & 48.389259 & 0.054823 & 17.69 & $-$19.43 \\
... & ... & ... & ... & ... & ... & ... & ... \\
\hline
\end{tabular}
    \tablefoot{The full table is available at the CDS. The columns correspond to: (1) group identification; (2) void identification; (3) CAVITY galaxy identification; (4) J2000.0 right ascension in degrees; (5) J2000.0 declination in degrees; (6) redshift; (7) apparent magnitude in the $r$-band; and (8) absolute magnitude in the $r$-band.}
\end{table*}

\begin{table*}
    \centering
    \caption{\label{tab:groups}Catalogue of groups in voids.}
\begin{tabular}{cccccccc}
\hline\hline
(1) & (2) & (3) & (4) & (5) & (6) & (7) & (8) \\
groupID & $\rm N_{gal}$ & $\rm \log(M_\bigstar)$ & $\rm L_r$ & $\rm \sigma_{v_r}$  & $\rm R_H$ & $\rm H_0 t_c$ & $\rm \log(M_{vir})$ \\
&   & (log(M$_\odot$)) & (10$^{10}$L$_\odot$) & (km\,s$^{-1}$) & (kpc) &  & (log(M$_\odot$)) \\
\hline\hline
1 & 2 & 10.58 & 2.74 & 50.45 & 233.15 & 0.59 & 12.41 \\
2 & 2 & 10.57 & 4.30 & 0.17\tablefootmark{a} & 63.52 & 46.60 & 6.92 \\
3 & 2 & 10.67 & 2.47 & 32.86 & 99.35 & 0.38 & 11.67 \\
4 & 2 & 10.26 & 1.80 & 31.37 & 393.49 & 1.59 & 12.23 \\
5 & 2 & 10.49 & 2.38 & 40.07 & 137.53 & 0.44 & 11.98 \\
6 & 2 & 10.39 & 1.80 & 0.17\tablefootmark{a} & 22.89 & 16.79 & 6.48 \\
7 & 2 & 10.56 & 2.49 & 16.53 & 337.93 & 2.59 & 11.60 \\
8 & 4 & 10.70 & 3.03 & 88.65 & 159.71 & 0.23 & 12.56 \\
9 & 2 & 10.47 & 2.43 & 9.12 & 252.53 & 3.52 & 10.96 \\
10 & 2 & 10.61 & 2.92 & 32.65 & 169.29 & 0.66 & 11.90 \\
... & ... & ... & ... & ... & ... & ... & ... \\
\hline
\end{tabular}
    \tablefoot{The full table is available at the CDS. The columns correspond to: (1) group identification; (2) group richness; (3) total stellar mass of the group in log(M$_\odot$); (4) total $r$-band luminosity of the group in 10$^{10}$L$_\odot$; (5) $\rm \sigma_{v_r}$, velocity dispersion in km\,s$^{-1}$; (6) $\rm R_H$, harmonic radius in kpc; (7) $\rm H_0 t_c$, crossing time; and (8) $\rm M_{vir}$, virial mass in log(M$_\odot$). \tablefoottext{a}{Note that groups with $\rm \sigma_{v_r}^2\,<\,10\, km\,s^{-1}$ have unreliable virial mass estimations.}}
\end{table*}

\subsection{Dynamical stage of galaxy groups} \label{sec:res_dynpar}

\begin{figure*}
\begin{center} 
\includegraphics[width=0.255\textwidth]{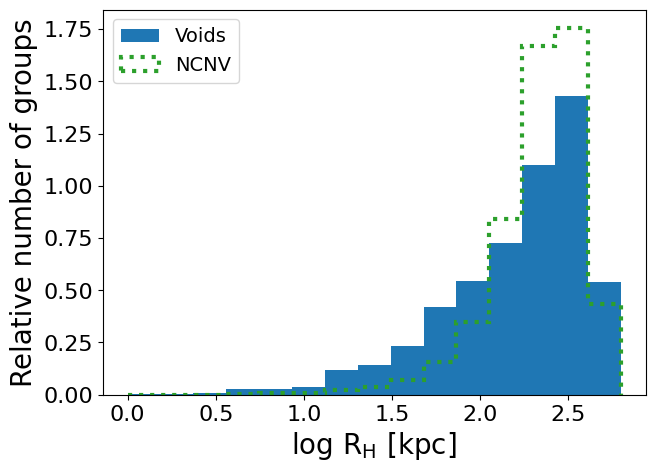}
\includegraphics[width=0.24\textwidth]{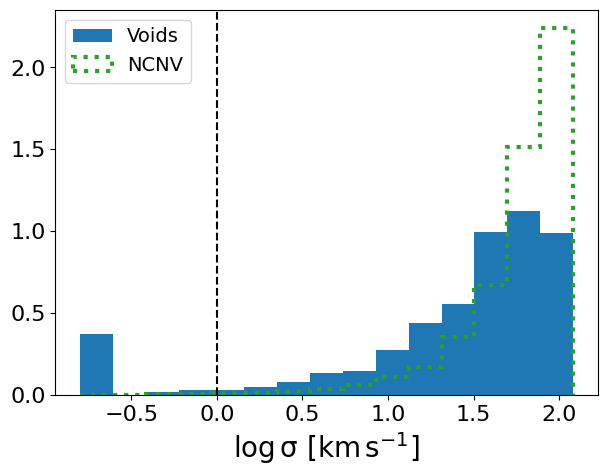}
\includegraphics[width=0.24\textwidth]{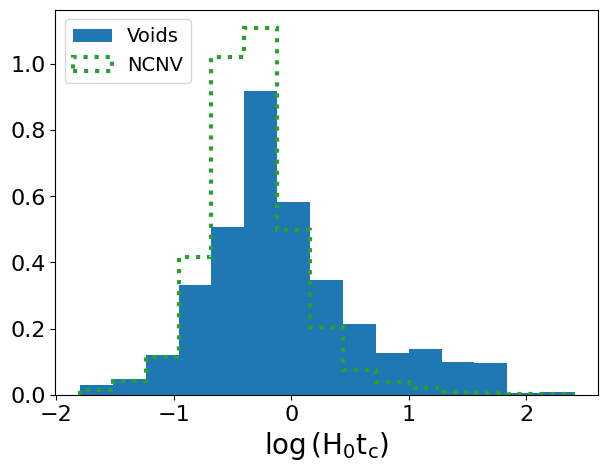}
\includegraphics[width=0.24\textwidth]{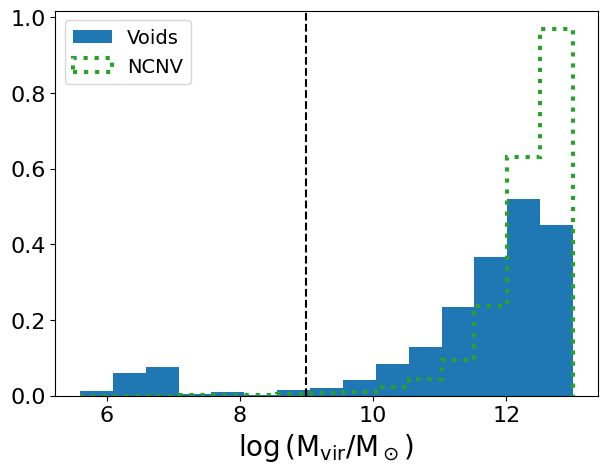}
\caption[]{Distribution of the dynamical parameters for the 1397 galaxy groups in the full sample. From left to right: effective radius of the groups, as $\rm log~R_H$; observed velocity dispersion, as $\rm log\, \sigma$; dimensionless crossing time, as $\rm log\, (H_0 t_c)$; and group virial mass, as $\rm log\, M_{vir}$ in solar units. The black dashed line in the distribution of the observed velocity dispersion delimits the groups with $\rm \sigma^2\,<\,10\, km\,s^{-1}$, and by consequence with $\rm log\, (M_{vir}/M_\odot)\,<\,9$, which are excluded of the analysis of the groups using the group virial mass. The distributions of the parameters for the control sample of NCNV groups are presented as green dotted open histograms.} \label{fig:dynpar} 
 \end{center}
\end{figure*}

Figure~\ref{fig:dynpar} shows the distribution of the dynamical parameters for the full sample of groups in voids. The median value of the dynamical parameters for the groups are $\rm R_H$\,=\,204$_{99}^{230}$\,kpc, $\rm \sigma_{v_r}$\,=\,43$_{21}^{72}$\,km\,s$^{-1}$, $\rm H_0 t_c$\,=\,0.66$_{0.34}^{1.81}$, and $\rm  \log(M_{vir}/M_\odot)$\,=\,12.2$_{11.5}^{12.6}$; where the upper and lower indices correspond to the 25$^{th}$ and 75$^{th}$ percentiles of the distribution, respectively.
In comparison, the median values of these parameters in the control sample of NCNV groups are $\rm R_H$\,=\,240$_{163}^{326}$\,kpc, $\rm \sigma_{v_r}$\,=\,73$_{48}^{94}$\,km\,s$^{-1}$, $\rm H_0 t_c$\,=\,0.44$_{0.26}^{0.73}$, and $\rm  \log(M_{vir}/M_\odot)$\,=\,12.5$_{12.2}^{12.8}$. This indicates that, in general, galaxy groups in voids are less massive, less compact, and less evolved than in denser large-scale environments.

\begin{figure*}
\begin{center} 
\includegraphics[width=0.45\textwidth]{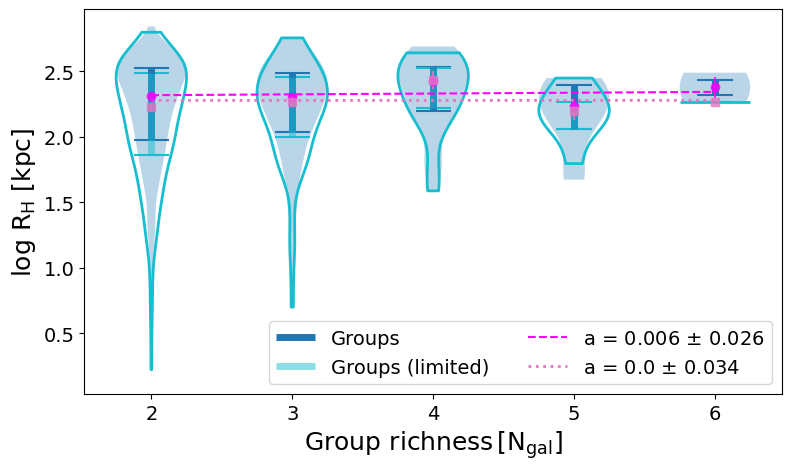}
\includegraphics[width=0.45\textwidth]{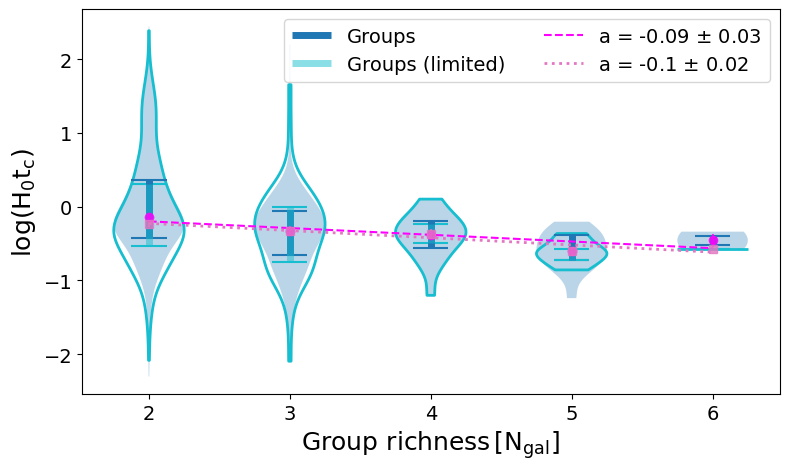}
\caption[]{Dynamical parameters of the galaxy groups with more than one member as a function of the group richness. The effective radius of the groups, as $\rm log~R_H$ is shown in the left panel, while the dimensionless crossing time, as $\rm log\, (H_0 t_c)$ is shown in the right panel. The observed velocity dispersion and group virial mass are not shown since their correlation with richness is expected, by definition (presented in equations~\ref{eq:eq2} and \ref{eq:eq4}). The distribution of the dynamical parameters for groups with same group richness is represented by violin plots (blue filled distributions considering the full sample, and cyan empty distributions for the limited sample). The inner box in each violin plot represents the interquartile range of the median and its 95\% of confidence intervals. Magenta circles and pink squares correspond to the median values in richness bins (considering the full sample or the limited sample, respectively) with their corresponding uncertainties, while dashed magenta line and dotted pink line correspond to the linear least-square fit to the binned data, respectively. The slopes of the linear fits, with their corresponding uncertainties, are indicted in the legend.} \label{fig:dynparNgal} 
\end{center}
\end{figure*}

Similar to Fig.~\ref{fig:voidprop}, in Fig.~\ref{fig:dynparNgal} we explore the dependence of the dynamical parameters for the groups in voids with respect to the group richness. The dispersion of the distributions of the dynamical parameters decreases with group richness, with galaxy pairs presenting the largest dispersion. The distributions for the limited sample are almost identical, indicating that these parameters are free of a redshift bias. In general, the median harmonic radius of groups does not vary with group richness, while the median radial velocity dispersion increases with group richness as is expected by definition (not shown). Consequently, the crossing time decreases with group richness while the virial mass increases. This trend suggests that groups become more compact and massive as the number of neighbours increases, as expected. When considering the control sample of NCNV groups, the median harmonic radius slightly decreases with group richness (slope a\,=\,$-$0.023\,$\pm$\,0.002), indicating that the groups are more compact with an increasing number of galaxies. The correlation of $\rm H_0 t_c$ with group richness is similar (slope a\,=\,$-$0.06\,$\pm$\,0.01). This indicates that galaxy groups in voids are less compact than in NCNV.

\subsection{Mass-to-light ratio of galaxy groups} \label{sec:res_ML}

\begin{figure}
\begin{center} 
\includegraphics[width=0.9\columnwidth]{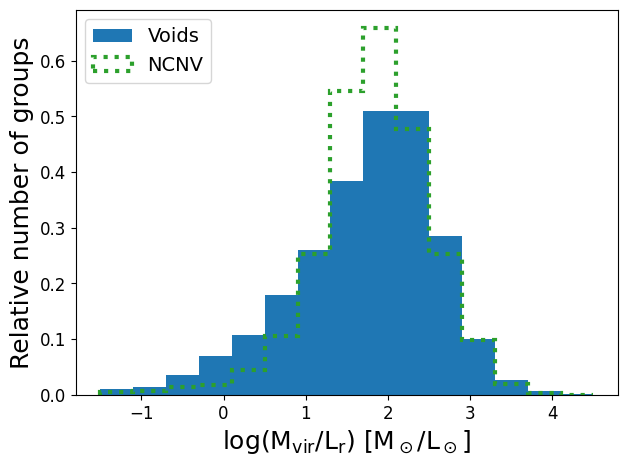}
\caption[]{Distribution of the mass-to-light ratio of the groups in the full sample with $\rm N_{gal}\,\geq\,2$. Only groups with $\rm log\, (M_{vir}/M_\odot)\,>\,9$ are considered in the analysis. The distribution for the control sample of NCNV groups is presented as green dotted open histogram.} \label{fig:ML} 
 \end{center}
\end{figure}

We used the total $r$-band luminosity, L$_r$, as calculated in Sect.~\ref{sec:ML}, and the virial mass from Sect.~\ref{sec:dynpar} to estimate the mass-to-light ratio ($\rm M_{vir}/L_r$) for the galaxy groups in voids. Note that groups with $\rm \sigma_{v_r}^2\,<\,10\, km\,s^{-1}$ (which occurs in groups composed of two galaxies), which have unreliable virial mass estimations, are excluded from the analysis. The distribution of the $\rm \log(M_{vir}/L_r)$, in units of $\rm \log(M_\odot/L_\odot)$, is shown in Fig.~\ref{fig:ML}. As expected, there is a wide range of values in the groups, with a median value of $\rm \log(M_{vir}/L_r)$\,=\,1.9$_{1.2}^{2.3}$, which corresponds to $\rm M_{vir}/L_r$\,$\sim$\,105\,$h$; where the upper and lower indices correspond to the 25$^{th}$ and 75$^{th}$ percentiles of the distribution, respectively. The results for the control sample of NCNV groups are similar, with slightly less dispersion at lower mass-to-light ratio ($\rm \log(M_{vir}/L_r)$\,=\,1.9$_{1.5}^{2.3}$).

\begin{figure}
\begin{center} 
\includegraphics[width=\columnwidth]{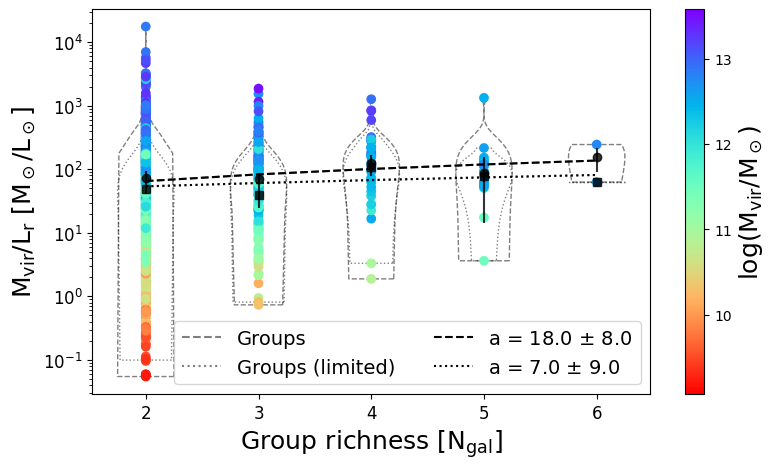}\\
\includegraphics[width=\columnwidth]{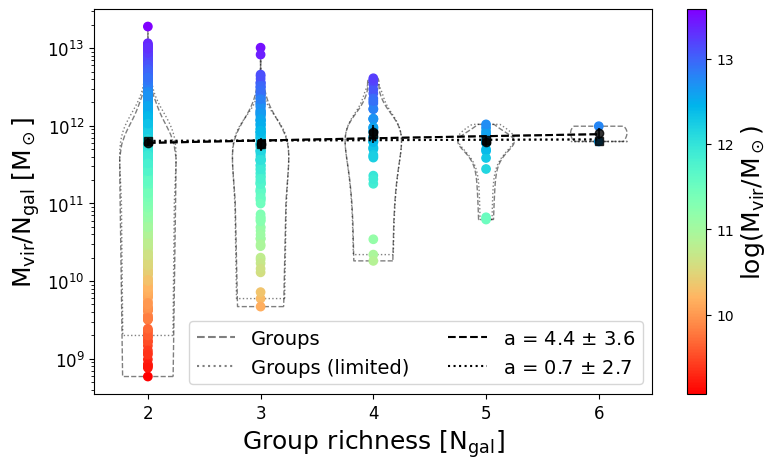}\\
\includegraphics[width=\columnwidth]{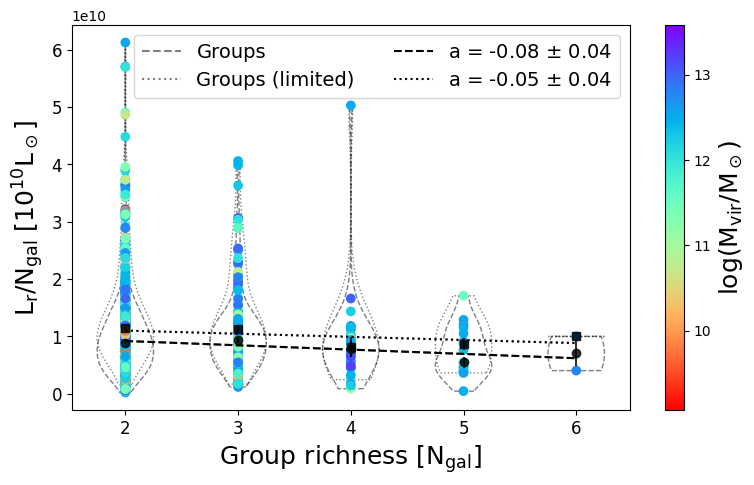}
\caption[]{Group mass-to-light ratio (top panel), mass-richness ratio (middle panel), and luminosity-richness ratio (bottom panel) as functions
of group richness. The distribution of these parameters for groups with same group richness is represented by violin plots (grey dashed-lines distributions considering the full sample, and grey dotted-lines distributions for the limited sample). Different virial group masses are indicated according to the colour bar at the right of each panel. Black circles correspond to the median values in richness bins with their corresponding uncertainties, while dashed black lines correspond to the linear least-square fits to the binned data. The slopes of the linear fits, with their corresponding uncertainties, are indicted in the legend. Only groups with $\rm log\, (M_{vir}/M_\odot)\,>\,9$ are considered in the analysis.} \label{fig:MvirLtot} 
 \end{center}
\end{figure}

In the top panel of Fig.~\ref{fig:MvirLtot}, we show how the $\rm M_{vir}/L_r$ varies with group richness, presenting a mild tendency to increase with the number of physically bound members, with a slope a\,=\,18\,$\pm$\,8 (a\,=\,7\,$\pm$\,9 for the limited sample). Additionally, we present the mass-richness ratio ($\rm M_{vir}/N_{gal}$) and luminosity-richness ratio ($\rm L_{r}/N_{gal}$) as functions of the group richness in the same figure. The $\rm M_{vir}/N_{gal}$ and $\rm L_{r}/N_{gal}$ show a very slight tendency to increase with richness (with large uncertainties), or is almost constant over the full richness range, respectively. The relation of the $\rm M_{vir}/N_{gal}$ with the group richness has a slope 4\,$\pm$\,3 (1\,$\pm$\,4 for the limited sample); while the $\rm L_{r}/N_{gal}$ with the group richness has a slope $-$0.08\,$\pm$\,0.04, compatible with the slope for the limited sample ($-$0.05\,$\pm$\,0.04), showing no trend with the group richness.  

\section{Discussion}  \label{sec:dis}


Voids are extremely underdense regions, with average density contrast of about $-$0.9 with diameters generally from 20 to 100\,Mpc \citep{2012MNRAS.421..926P}. This means that void galaxies are very sparse, which very commonly leads to the misconception of void galaxies as isolated galaxies. However, as introduced in Sect.~\ref{sec:intro}, many works have shown that there are substructures within voids. In this work, we have identified a sample of physically bound groups within voids. Considering the group richness of the full sample, 59\% of void galaxies are singlets, from which only 7\% are potential isolated galaxies considering the observational constraints (as discussed in Sect.~\ref{sec:pipeline}), while 12\% of the void galaxies belong to physically bound groups (3040 galaxies). The remaining void galaxies (29\%) are connected to the large-scale structure within the voids (i.e., they have at least one neighbour with $\rm \Delta v\,\leq\,\pm~500\,km\,s^{-1}$ at 1\,Mpc projected distance), but are not physically bound with it. 

\begin{figure*}
\begin{center} 
\includegraphics[width=0.49\linewidth, trim={0 1cm 0 0},clip]{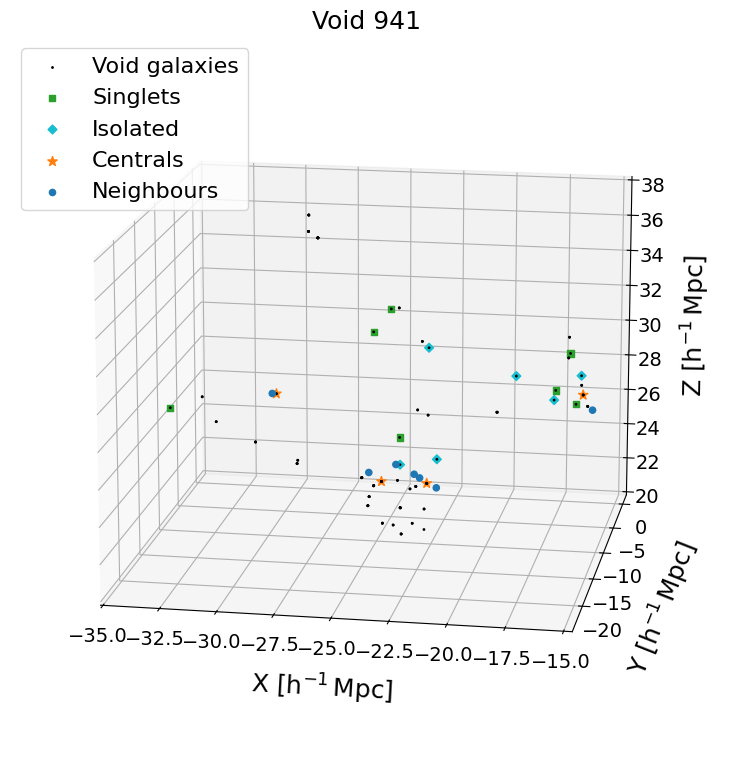}
\includegraphics[width=0.49\linewidth, trim={0 1cm 0 0},clip]{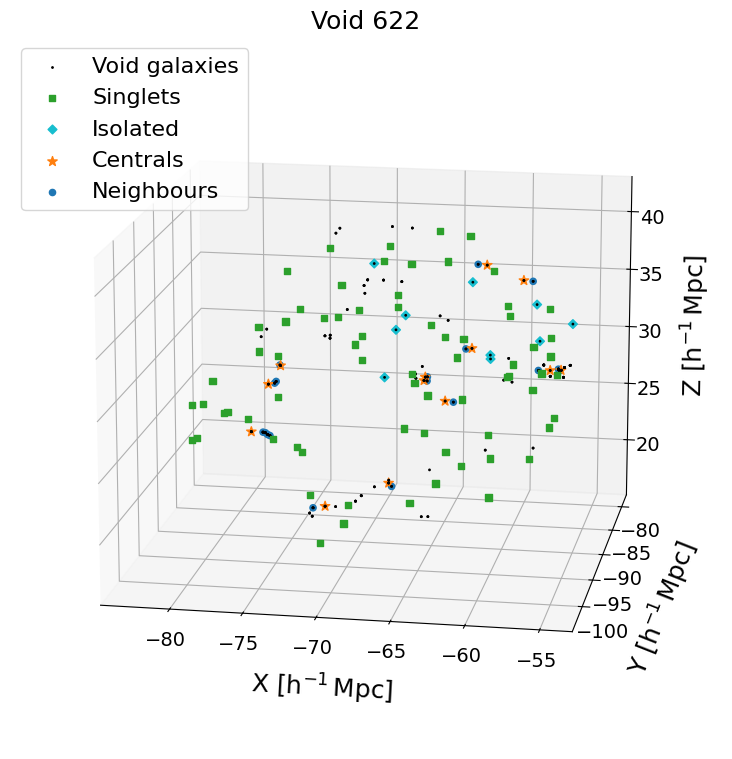}
\caption[]{3D visualisation of the spatial distribution of galaxies within two voids in the CAVITY sample (Void 941 and Void 622, in the left and right panels, respectively) as an example of the distribution of groups within voids of similar size ($\rm R_{void}$\,$\sim$\,13\,h$^{-1}$\,Mpc) but different number of galaxies (53 and 154 void galaxies, respectively). Void galaxies (black dots) act as tracers of the substructure of the voids. Of them, singlets (void galaxies with no neighbours $\rm \Delta v\,\leq\,\pm~500\,km\,s^{-1}$ at 1\,Mpc projected distance) are marked in green, from which potential isolated galaxies (singlets with $\rm m_{r}\,\leq\,$15.7\,mag) are marked in cyan, while central galaxies of groups with two or more members are marked in orange, with their corresponding physically bound members in blue, as indicated in the legend.} \label{fig:void3D} 
 \end{center}
\end{figure*}

We explored if group richness has any dependency with the properties of the voids, as void richness (number of galaxies in each void), void size, and the galaxy number-density of galaxies in a void (defined as the number of galaxies inhabiting a void divided by the total volume of the void). It is reasonable to assume that the densest groups are preferentially found in denser voids, as galaxies are more likely to become gravitationally bound to nearby neighbours and form groups in deeper potential wells. However, as shown in Fig.~\ref{fig:voidprop}, after correcting by redshift bias, we found that group richness is independent of the density of the voids. This means that we can find galaxy groups, and therefore substructures, in any void in the local Universe, with densities from 0.001 to 0.017 gal/(h$^{-1}$Mpc)$^3$. For reference, in Fig.~\ref{fig:void3D} we show the 3 dimensional visualisation of two voids of similar size ($\rm R_{void}$\,$\sim$\,13\,h$^{-1}$\,Mpc), but different number of galaxies: Void 941 ($\rm \rho_N$\,=\,0.005\,gal/(h$^{-1}$Mpc)$^3$ at $z$\,=\,0.01) in the left panel, with 53 void galaxies, and Void 622 ($\rm \rho_N$\,=\,0.014\,gal/(h$^{-1}$Mpc)$^3$ at $z$\,=\,0.04) in the right panel, with 154 void galaxies, thus about 3 times denser than Void 941. These two voids are included in the CAVITY parent sample. As expected, the more galaxies there are in a void, the greater the number of groups, however, the richness of the groups does not depend on how dense the void is.

The analysis of the values of the median harmonic radius, radial velocity dispersion, and crossing time found in this work, indicates that galaxy groups within voids can be considered as loose groups \citep{1987ApJ...321..622M}. Even the densest groups in voids are more loose than similar richness groups outside voids. Loose groups are characterised by presenting low velocity dispersions \citep[less than 200\,km\,s$^{-1}$,][]{2009AJ....138..338S} and galaxy separations by distances many times their own size (few hundred kpc), forming sparse structures, independently of the group richness. These systems are low-density, typically containing spiral-rich members, and represent the most common environment for galaxies (examples include the Leo Triplet or the Local Group), which can be also found in the vicinity of galaxy clusters \citep{2003A&A...401..851E}. The low dispersion of the velocity of the void galaxies might favour galaxy interactions and mergers within the groups, however, unlike in compact groups (median $\rm H_0 t_c$\,=\,0.016), crossing times values in the void groups are much larger (median $\rm H_0 t_c$\,=\,0.66 for the full sample), even larger than in the control sample of NCNV groups (median $\rm H_0 t_c$\,=\,0.44) and in isolated triplets \citep[median $\rm H_0 t_c$\,=\,0.38,][]{2023A&A...670A..63V} for the groups with higher richness, which indicates the early stage of evolution of group galaxies in voids. On the other hand, since group in voids are inherently loose and of low density, long crossing times and a lack of virialisation\footnote{Groups are supposed to be virialised if their crossing times are much smaller than the Hubble time, typically 0.1-0.5\,H$^{-1}_0$ \citep{2017MNRAS.471....2P}.} could be expected even without an early stage. In this scenario, galaxy groups in voids could be stable structures or in prolonged pseudo-equilibrium, due to the absence of frequent mergers. In other words, non-virialisation in void groups could be a permanent state due to global gravitational suppression, not a transitory one. However, the rate of mergers in voids at present Universe is comparable to the rate in filaments and walls ($\sim$\,3\,\%, Vásquez-Bustos et al. in preparation). Therefore interactions are expected to occur within the groups, even if the void galaxies tend to experience mergers at later times than in denser environments \citep[as found from simulations,][]{2024MNRAS.528.2822R}, which supports the first scenario where groups in voids are in an early evolutionary stage. This is also in agreement with recent observations which found that, more evolved groups are found in dense structures (clusters and superclusters), while poor groups reside everywhere in the cosmic web, where the less luminous groups at same richness (therefore less evolved) are found in low-density environments, as voids \citep{2024A&A...681A..91E}.

The median mass-to-light ratio for galaxy groups in voids ($\rm M_{vir}/L_r$\,$\sim$\,105\,$h$) is almost the double than the value reported for compact groups \citep[$\rm M_{vir}/L_r$\,$=$\,50\,$h$,][]{1992ApJ...399..353H}. This might suggest a larger dark matter fraction in void groups, in agreement with recent simulations \citep{2020MNRAS.493..899H,2020MNRAS.491.5747M,2022MNRAS.517..712R,2024MNRAS.528.2822R}. Unfortunately, our derived virial mass estimates for the groups (based on the group velocity dispersion, see Eq.~\ref{eq:eq4}), have large uncertainties owing to the small number of galaxies per group. In fact, the median mass-to-light ratio for the control sample of NCNV groups with same richness is similar to the groups in voids. Therefore, no strong conclusions can be drawn in this regard.
A dedicated study focused on identifying faint satellites in these groups would help to obtain more robust results. In this work, we limited the study to the analysis of the mass-to-light ratio as function of group richness, to compare with the observed trends of groups in denser environments. 

Figure~\ref{fig:MvirLtot} shows the $\rm M_{vir}/L_r$, $\rm M_{vir}/N_{gal}$, and $\rm L_{r}/N_{gal}$ as a function of group richness. The trends found for the mass-to-light ratio and the $\rm L_{r}/N_{gal}$ (after correcting for redshift bias) are in agreement with the trends reported in \citet{2017A&A...602A.100T}, which only consider groups with richness greater than $\rm N_{gal}$\,=\,6. When extending to low richness groups, the mass-to-light ratio increases slightly with group richness while the $\rm L_{r}/N_{gal}$ relation is almost flat. This flat luminosity-richness relation has been also found in previous works based on galaxy clusters in the SDSS \citep{2007A&A...464..451P,2009ApJS..183..197W}. As also pointed out by \citet{2017A&A...602A.100T}, the mass-to-light ratio decreases with group richness when considering groups with similar mass, even with the wide range of $\rm M_{vir}/L_r$ values that we found for low richness groups. However, we found that the $\rm M_{vir}/N_{gal}$ relation increases much slightly with group richness than reported in observational studies on denser groups and galaxy clusters \citep{2017A&A...602A.100T,2002ApJ...569..101M,2003AJ....126.1677P,2009ApJS..183..197W}. From simulations, the $\rm M_{vir}/N_{gal}$ relation (so-called halo occupation distribution) can be described by a power law $\rm N_{gal}\,\propto\,M^{\beta}_{vir}$, with $\beta$ presenting expected values less than 1 \citep{2001ApJ...550L.129W}. With our data, we find $\beta$\,=\,0.71\,$\pm$\,0.13 ($\beta$\,=\,0.98\,$\pm$\,0.37 for the limited sample). Although with large uncertainties in the $\rm M_{vir}$ values, the $\rm M_{vir}/N_{gal}$ relation shows the expected behaviour. The very slight increase with group richness may be due to the early stage of evolution of galaxy groups in voids and the sparse distance between the galaxies. Therefore the groups, although physically bound, they may still not be virialised. According to \citet{2026MNRAS.545f2069W}, who studied observed red luminous galaxies in contrast to simulations, the clustering of these type of groups is directly connected to the underlying matter density field, which provides key information about non-linear structure formation and the galaxy–halo connection.

\section{Summary and conclusions}  \label{sec:con}

In this work we have identified a sample of galaxy groups in 170 cosmic voids in the local Universe ($z$\,<\,0.08), taking into account the peculiarities of these vast and empty structures. For this, each void has been analysed separately, considering only the galaxies that reside in it, and we applied a FoF-like group finder algorithm, ensuring a certain degree of gravitational bounding among group members. The sample of void galaxies that we use in this study is based on the well-defined sample of void galaxies from \citet{2012MNRAS.421..926P} (also used in the CAVITY legacy project). The catalogue of groups is composed of 1367 physically bound groups (with a total of 3040 galaxies), and 14672 galaxy singlets. We used the same group finder algorithm to identify a control sample of groups considering galaxies not part of either clusters nor voids (NCNV).

To consider the dynamical stage of the groups we used the following parameters: harmonic radius ($\rm R_H$), radial velocity dispersion ($\rm \sigma_{v_r}^2$), dimensionless crossing time ($\rm H_0 t_c$), and group virial mass ($\rm  M_{vir}$), as explained in Sect.~\ref{sec:dynpar}. We also used the total optical ($r$-band) luminosity, L$_r$, to estimate the mass-to-light ratio ($\rm M_{vir}/L_r$) for the galaxy groups in voids and in NCNV. We studied the relations of the group richness with void properties, the dynamical parameters, and the mass-to-light ratio. Our main findings are as follows:

\begin{itemize}
    \item Galaxy groups can be found in any void in the local Universe. These are evenly distributed within voids. The more galaxies there are in a void, the greater the number of groups, however, the group richness is independent of the density of the voids.
    \item Most of the galaxies in voids are singlet galaxies (59\%), from which only a few percent (7\%) are potentially isolated galaxies, considering the observational constraints. In contrast, most of the galaxies in a control sample in NCNV are in groups (60\%). The densest groups in voids are composed of six galaxies, therefore, voids generally contain small groups, in comparison to denser structures such as filaments, walls, and galaxy clusters. 
    \item Considering their dynamical properties, galaxy groups within voids are typically loose groups, in an early stage of their evolution. The groups are more compact and massive with increasing number of galaxies, favouring galaxy interactions between their members. 
    \item The work presented here has allowed us to extend the mass-to-light ratio, mass-richness ratio, and luminosity-richness ratio as functions of richness relations to lower ambient density values, confirming the trends observed in denser groups and clusters. However, the mass-richness relation is much flatter, which might be due to the early stage of evolution of galaxy groups in voids and the sparse distance between their galaxies. This result suggests that galaxy groups in voids, although physically bound, may still not be virialised.
\end{itemize}

The catalogue of group galaxies, with galaxy positions, redshifts, and optical magnitudes, as well as the catalogue of groups with group richness, total stellar mass, total optical luminosity, and dynamical parameters are publicly available to the scientific community. The data structure and first lines in our catalogues are described in Tables~\ref{tab:galaxies} and \ref{tab:groups}.

\begin{acknowledgements}
The CAVITY project acknowledges financial support by the research projects AYA2017-84897-P, PID2020-113689GB-I00, PID2020-114414GB-I00, and PID2023-149578NB-I00 funded by the Spanish Ministry of Science and Innovation (MCIN/AEI/10.13039/501100011033) and by FEDER/UE; the project A-FQM-510-UGR20 funded by FEDER/Junta de Andalucía-Consejería de Transformación Económica, Industria, Conocimiento y Universidades/Proyecto; by the grants P20-00334 and FQM108, funded by Junta de Andalucía; and by Consejería de Universidad, Investigación e Innovación (Junta de Andalucía) and Gobierno de España and European Union NextGenerationEU through grant AST22\_4.4. 
We also acknowledge the research project PID2023-150178NB-I00 by Consejería de Universidad, Investigación e Innovación and Gobierno de España.
MAF and PVB acknowledges support from the Emergia program (EMERGIA20\_38888) from Consejer\'ia de Universidad, Investigaci\'on e Innovaci\'on de la Junta de Andaluc\'ia.
GTR acknowledges financial support from the research project PRE2021-098736, funded by MCIN/AEI/10.13039/501100011033 and the FSE+.
AC, RGB, and RGD acknowledge financial support from the Severo Ochoa grant CEX2021-001131-S funded by MCIN/AEI/ 10.13039/501100011033, and PID2022-141755NB-I00.
This work has been supported by the Agencia Estatal de Investigación Española (AEI; grant PID2022-138855NB-C33), by the Ministerio de Ciencia e Innovación (MCIN) within the Plan de Recuperación, Transformación y Resiliencia del Gobierno de España through the project ASFAE/2022/001, with funding from European Union NextGenerationEU (PRTR-C17.I1), and by the Generalitat Valenciana (grant CIPROM/2022/49).
IMC acknowledge support from ANID programme FONDECYT Postdoctorado 3230653 and ANID, BASAL, FB210003.
JR acknowledges financial support from Plan Propio de Investigación 2025 submodalidad 2.3 of the University of Cordoba.
SS acknowledges the support from the Alexander von Humboldt Foundation.\\
Funding for SDSS-III has been provided by the Alfred P. Sloan Foundation, the Participating Institutions, the National Science Foundation, and the U.S. Department of Energy Office of Science. The SDSS-III web site is http://www.sdss3.org/. SDSS-III is managed by the Astrophysical Research Consortium for the Participating Institutions of the SDSS-III Collaboration including the University of Arizona, the Brazilian Participation Group, Brookhaven National Laboratory, Carnegie Mellon University, University of Florida, the French Participation Group, the German Participation Group, Harvard University, the Instituto de Astrofisica de Canarias, the Michigan State/Notre Dame/JINA Participation Group, Johns Hopkins University, Lawrence Berkeley National Laboratory, Max Planck Institute for Astrophysics, Max Planck Institute for Extraterrestrial Physics, New Mexico State University, New York University, Ohio State University, Pennsylvania State University, University of Portsmouth, Princeton University, the Spanish Participation Group, University of Tokyo, University of Utah, Vanderbilt University, University of Virginia, University of Washington, and Yale University.\\
This research made use of Astropy, a community-developed core Python (http://www.python.org) package for Astronomy \citep{astropy:2013, astropy:2018, astropy:2022}; ipython \citep{PER-GRA:2007}; matplotlib \citep{Hunter:2007}; SciPy, a collection of open source software for scientific computing in Python \citep{2020SciPy-NMeth}; pandas, an open source data analysis and manipulation tool \citep{reback2020pandas, mckinney-proc-scipy-2010}; and NumPy, a structure for efficient numerical computation \citep{2011CSE....13b..22V}.
\end{acknowledgements}
\bibliographystyle{aa} 
\bibliography{refs} 

@ARTICLE{2009ApJS..182..543A,
       author = {{Abazajian}, Kevork N. and {Adelman-McCarthy}, Jennifer K. and {Ag{\"u}eros}},
        title = "{The Seventh Data Release of the Sloan Digital Sky Survey}",
      journal = {\apjs},
         year = 2009,
        month = jun,
       volume = {182},
       number = {2},
        pages = {543-558},
          doi = {10.1088/0067-0049/182/2/543},
archivePrefix = {arXiv},
       eprint = {0812.0649},
 primaryClass = {astro-ph},
       adsurl = {https://ui.adsabs.harvard.edu/abs/2009ApJS..182..543A}
}

@ARTICLE{1989ApJS...70....1A,
       author = {{Abell}, George O. and {Corwin}, Jr., Harold G. and {Olowin}, Ronald P.},
        title = "{A Catalog of Rich Clusters of Galaxies}",
      journal = {\apjs},
         year = 1989,
        month = may,
       volume = {70},
        pages = {1},
          doi = {10.1086/191333},
       adsurl = {https://ui.adsabs.harvard.edu/abs/1989ApJS...70....1A}
}

@ARTICLE{2015ApJS..219...12A,
       author = {{Alam}, Shadab and {Albareti}, Franco D. and {Allende Prieto}, Carlos and {Anders}, F. and {Anderson}, Scott F. and {Anderton}, Timothy and {Andrews}, Brett H. and {Armengaud}, Eric and {Aubourg}, {\'E}ric and {Bailey}, Stephen and {Basu}, Sarbani and {Bautista}, Julian E. and {Beaton}, Rachael L. and {Beers}, Timothy C. and {Bender}, Chad F. and {Berlind}, Andreas A. and {Beutler}, Florian and {Bhardwaj}, Vaishali and {Bird}, Jonathan C. and {Bizyaev}, Dmitry and {Blake}, Cullen H. and {Blanton}, Michael R. and {Blomqvist}, Michael and {Bochanski}, John J. and {Bolton}, Adam S. and {Bovy}, Jo and {Shelden Bradley}, A. and {Brandt}, W.~N. and {Brauer}, D.~E. and {Brinkmann}, J. and {Brown}, Peter J. and {Brownstein}, Joel R. and {Burden}, Angela and {Burtin}, Etienne and {Busca}, Nicol{\'a}s G. and {Cai}, Zheng and {Capozzi}, Diego and {Carnero Rosell}, Aurelio and {Carr}, Michael A. and {Carrera}, Ricardo and {Chambers}, K.~C. and {Chaplin}, William James and {Chen}, Yen-Chi and {Chiappini}, Cristina and {Chojnowski}, S. Drew and {Chuang}, Chia-Hsun and {Clerc}, Nicolas and {Comparat}, Johan and {Covey}, Kevin and {Croft}, Rupert A.~C. and {Cuesta}, Antonio J. and {Cunha}, Katia and {da Costa}, Luiz N. and {Da Rio}, Nicola and {Davenport}, James R.~A. and {Dawson}, Kyle S. and {De Lee}, Nathan and {Delubac}, Timoth{\'e}e and {Deshpande}, Rohit and {Dhital}, Saurav and {Dutra-Ferreira}, Let{\'\i}cia and {Dwelly}, Tom and {Ealet}, Anne and {Ebelke}, Garrett L. and {Edmondson}, Edward M. and {Eisenstein}, Daniel J. and {Ellsworth}, Tristan and {Elsworth}, Yvonne and {Epstein}, Courtney R. and {Eracleous}, Michael and {Escoffier}, Stephanie and {Esposito}, Massimiliano and {Evans}, Michael L. and {Fan}, Xiaohui and {Fern{\'a}ndez-Alvar}, Emma and {Feuillet}, Diane and {Filiz Ak}, Nurten and {Finley}, Hayley and {Finoguenov}, Alexis and {Flaherty}, Kevin and {Fleming}, Scott W. and {Font-Ribera}, Andreu and {Foster}, Jonathan and {Frinchaboy}, Peter M. and {Galbraith-Frew}, J.~G. and {Garc{\'\i}a}, Rafael A. and {Garc{\'\i}a-Hern{\'a}ndez}, D.~A. and {Garc{\'\i}a P{\'e}rez}, Ana E. and {Gaulme}, Patrick and {Ge}, Jian and {G{\'e}nova-Santos}, R. and {Georgakakis}, A. and {Ghezzi}, Luan and {Gillespie}, Bruce A. and {Girardi}, L{\'e}o and {Goddard}, Daniel and {Gontcho}, Satya Gontcho A. and {Gonz{\'a}lez Hern{\'a}ndez}, Jonay I. and {Grebel}, Eva K. and {Green}, Paul J. and {Grieb}, Jan Niklas and {Grieves}, Nolan and {Gunn}, James E. and {Guo}, Hong and {Harding}, Paul and {Hasselquist}, Sten and {Hawley}, Suzanne L. and {Hayden}, Michael and {Hearty}, Fred R. and {Hekker}, Saskia and {Ho}, Shirley and {Hogg}, David W. and {Holley-Bockelmann}, Kelly and {Holtzman}, Jon A. and {Honscheid}, Klaus and {Huber}, Daniel and {Huehnerhoff}, Joseph and {Ivans}, Inese I. and {Jiang}, Linhua and {Johnson}, Jennifer A. and {Kinemuchi}, Karen and {Kirkby}, David and {Kitaura}, Francisco and {Klaene}, Mark A. and {Knapp}, Gillian R. and {Kneib}, Jean-Paul and {Koenig}, Xavier P. and {Lam}, Charles R. and {Lan}, Ting-Wen and {Lang}, Dustin and {Laurent}, Pierre and {Le Goff}, Jean-Marc and {Leauthaud}, Alexie and {Lee}, Khee-Gan and {Lee}, Young Sun and {Licquia}, Timothy C. and {Liu}, Jian and {Long}, Daniel C. and {L{\'o}pez-Corredoira}, Mart{\'\i}n and {Lorenzo-Oliveira}, Diego and {Lucatello}, Sara and {Lundgren}, Britt and {Lupton}, Robert H. and {Mack}, III, Claude E. and {Mahadevan}, Suvrath and {Maia}, Marcio A.~G. and {Majewski}, Steven R. and {Malanushenko}, Elena and {Malanushenko}, Viktor and {Manchado}, A. and {Manera}, Marc and {Mao}, Qingqing and {Maraston}, Claudia and {Marchwinski}, Robert C. and {Margala}, Daniel and {Martell}, Sarah L. and {Martig}, Marie and {Masters}, Karen L. and {Mathur}, Savita and {McBride}, Cameron K. and {McGehee}, Peregrine M. and {McGreer}, Ian D. and {McMahon}, Richard G. and {M{\'e}nard}, Brice and {Menzel}, Marie-Luise and {Merloni}, Andrea and {M{\'e}sz{\'a}ros}, Szabolcs and {Miller}, Adam A. and {Miralda-Escud{\'e}}, Jordi and {Miyatake}, Hironao and {Montero-Dorta}, Antonio D. and {More}, Surhud and {Morganson}, Eric and {Morice-Atkinson}, Xan and {Morrison}, Heather L. and {Mosser}, Ben{\^o}it and {Muna}, Demitri and {Myers}, Adam D. and {Nandra}, Kirpal and {Newman}, Jeffrey A. and {Neyrinck}, Mark and {Nguyen}, Duy Cuong and {Nichol}, Robert C. and {Nidever}, David L. and {Noterdaeme}, Pasquier and {Nuza}, Sebasti{\'a}n E. and {O'Connell}, Julia E. and {O'Connell}, Robert W. and {O'Connell}, Ross and {Ogando}, Ricardo L.~C. and {Olmstead}, Matthew D. and {Oravetz}, Audrey E. and {Oravetz}, Daniel J. and {Osumi}, Keisuke and {Owen}, Russell and {Padgett}, Deborah L. and {Padmanabhan}, Nikhil and {Paegert}, Martin and {Palanque-Delabrouille}, Nathalie and {Pan}, Kaike},
        title = "{The Eleventh and Twelfth Data Releases of the Sloan Digital Sky Survey: Final Data from SDSS-III}",
      journal = {\apjs},
         year = 2015,
        month = jul,
       volume = {219},
       number = {1},
          eid = {12},
        pages = {12},
          doi = {10.1088/0067-0049/219/1/12},
archivePrefix = {arXiv},
       eprint = {1501.00963},
 primaryClass = {astro-ph.IM},
       adsurl = {https://ui.adsabs.harvard.edu/abs/2015ApJS..219...12A}
}

@ARTICLE{2026arXiv260205117A,
       author = {{Alfaro}, Ignacio G. and {Montero-Dorta}, Antonio D. and {Bustillos}, Jorge F. and {Paz}, Dante J. and {Ruiz}, Andr{\'e}s N. and {Balaguera-Antol{\'\i}nez}, Andr{\'e}s and {Sheth}, Ravi K. and {Rodriguez}, Facundo and {Soto-Su{\'a}rez}, Constanza A.},
        title = "{The Galaxy Bias Profile of Cosmic Voids:A Comparison of Void Finders}",
      journal = {arXiv e-prints},
         year = 2026,
        month = feb,
          eid = {arXiv:2602.05117},
        pages = {arXiv:2602.05117},
          doi = {10.48550/arXiv.2602.05117},
archivePrefix = {arXiv},
       eprint = {2602.05117},
 primaryClass = {astro-ph.CO},
       adsurl = {https://ui.adsabs.harvard.edu/abs/2026arXiv260205117A}
}

@ARTICLE{2022A&A...665A..44A,
       author = {{Alfaro}, Ignacio G. and {Rodriguez}, Facundo and {Ruiz}, Andr{\'e}s N. and {Luparello}, Heliana E. and {Lambas}, Diego Garcia},
        title = "{How galaxies populate halos in extreme density environments: An analysis of the halo occupation distribution in SDSS}",
      journal = {\aap},
         year = 2022,
        month = sep,
       volume = {665},
          eid = {A44},
        pages = {A44},
          doi = {10.1051/0004-6361/202243542},
archivePrefix = {arXiv},
       eprint = {2203.07526},
 primaryClass = {astro-ph.CO},
       adsurl = {https://ui.adsabs.harvard.edu/abs/2022A&A...665A..44A}
}

@ARTICLE{2020A&A...638A..60A,
       author = {{Alfaro}, Ignacio G. and {Rodriguez}, Facundo and {Ruiz}, Andr{\'e}s N. and {Lambas}, Diego Garcia},
        title = "{How galaxies populate haloes in very low-density environments. An analysis of the halo occupation distribution in cosmic voids}",
      journal = {\aap},
         year = 2020,
        month = jun,
       volume = {638},
          eid = {A60},
        pages = {A60},
          doi = {10.1051/0004-6361/201937431},
archivePrefix = {arXiv},
       eprint = {2003.06255},
 primaryClass = {astro-ph.CO},
       adsurl = {https://ui.adsabs.harvard.edu/abs/2020A&A...638A..60A}
}

@ARTICLE{2014MNRAS.438..177A,
       author = {{Alpaslan}, Mehmet and {Robotham}, Aaron S.~G. and {Driver}, Simon and {Norberg}, Peder and {Baldry}, Ivan and {Bauer}, Amanda E. and {Bland-Hawthorn}, Joss and {Brown}, Michael and {Cluver}, Michelle and {Colless}, Matthew and {Foster}, Caroline and {Hopkins}, Andrew and {Van Kampen}, Eelco and {Kelvin}, Lee and {Lara-Lopez}, Maritza A. and {Liske}, Jochen and {Lopez-Sanchez}, Angel R. and {Loveday}, Jon and {McNaught-Roberts}, Tamsyn and {Merson}, Alexander and {Pimbblet}, Kevin},
        title = "{Galaxy And Mass Assembly (GAMA): the large-scale structure of galaxies and comparison to mock universes}",
      journal = {\mnras},
         year = 2014,
        month = feb,
       volume = {438},
       number = {1},
        pages = {177-194},
          doi = {10.1093/mnras/stt2136},
archivePrefix = {arXiv},
       eprint = {1311.1211},
 primaryClass = {astro-ph.CO},
       adsurl = {https://ui.adsabs.harvard.edu/abs/2014MNRAS.438..177A}
}

@ARTICLE{2014MNRAS.440L.106A,
       author = {{Alpaslan}, M. and {Robotham}, A.~S.~G. and {Obreschkow}, D. and {Penny}, S. and {Driver}, S. and {Norberg}, P. and {Brough}, S. and {Brown}, M. and {Cluver}, M. and {Holwerda}, B. and {Hopkins}, A.~M. and {van Kampen}, E. and {Kelvin}, L.~S. and {Lara-Lopez}, M.~A. and {Liske}, J. and {Loveday}, J. and {Mahajan}, S. and {Pimbblet}, K.},
        title = "{Galaxy and Mass Assembly (GAMA): fine filaments of galaxies detected within voids.}",
      journal = {\mnras},
         year = 2014,
        month = may,
       volume = {440},
        pages = {L106-L110},
          doi = {10.1093/mnrasl/slu019},
archivePrefix = {arXiv},
       eprint = {1401.7331},
 primaryClass = {astro-ph.CO},
       adsurl = {https://ui.adsabs.harvard.edu/abs/2014MNRAS.440L.106A}
}

@ARTICLE{2010MNRAS.404L..89A,
       author = {{Aragon-Calvo}, M.~A. and {van de Weygaert}, R. and {Araya-Melo}, P.~A. and {Platen}, E. and {Szalay}, A.~S.},
        title = "{Unfolding the hierarchy of voids}",
      journal = {\mnras},
         year = 2010,
        month = may,
       volume = {404},
       number = {1},
        pages = {L89-L93},
          doi = {10.1111/j.1745-3933.2010.00841.x},
archivePrefix = {arXiv},
       eprint = {1002.1503},
 primaryClass = {astro-ph.CO},
       adsurl = {https://ui.adsabs.harvard.edu/abs/2010MNRAS.404L..89A}
}

@ARTICLE{2024A&A...692A.258A,
       author = {{Argudo-Fern{\'a}ndez}, M. and {G{\'o}mez Hern{\'a}ndez}, C. and {Verley}, S. and {Zurita}, A. and {Duarte Puertas}, S. and {Bl{\'a}zquez Calero}, G. and {Dom{\'\i}nguez-G{\'o}mez}, J. and {Espada}, D. and {Florido}, E. and {P{\'e}rez}, I. and {S{\'a}nchez-Menguiano}, L.},
        title = "{Morphologies of galaxies within voids}",
      journal = {\aap},
         year = 2024,
        month = dec,
       volume = {692},
          eid = {A258},
        pages = {A258},
          doi = {10.1051/0004-6361/202450809},
archivePrefix = {arXiv},
       eprint = {2411.02129},
 primaryClass = {astro-ph.GA},
       adsurl = {https://ui.adsabs.harvard.edu/abs/2024A&A...692A.258A}
}

@ARTICLE{2015A&A...578A.110A,
       author = {{Argudo-Fern{\'a}ndez}, M. and {Verley}, S. and {Bergond}, G. and {Duarte Puertas}, S. and {Ramos Carmona}, E. and {Sabater}, J. and {Fern{\'a}ndez Lorenzo}, M. and {Espada}, D. and {Sulentic}, J. and {Ruiz}, J.~E. and {Leon}, S.},
        title = "{Catalogues of isolated galaxies, isolated pairs, and isolated triplets in the local Universe}",
      journal = {\aap},
         year = 2015,
        month = jun,
       volume = {578},
          eid = {A110},
        pages = {A110},
          doi = {10.1051/0004-6361/201526016},
archivePrefix = {arXiv},
       eprint = {1504.00117},
 primaryClass = {astro-ph.GA},
       adsurl = {https://ui.adsabs.harvard.edu/abs/2015A&A...578A.110A}
}

@article{astropy:2013,
Adsurl = {http://adsabs.harvard.edu/abs/2013A%26A...558A..33A},
Archiveprefix = {arXiv},
Author = {{Astropy Collaboration} and {Robitaille}, T.~P. and {Tollerud}, E.~J. and {Greenfield}, P. and {Droettboom}, M. and {Bray}, E. and {Aldcroft}, T. and {Davis}, M. and {Ginsburg}, A. and {Price-Whelan}, A.~M. and {Kerzendorf}, W.~E. and {Conley}, A. and {Crighton}, N. and {Barbary}, K. and {Muna}, D. and {Ferguson}, H. and {Grollier}, F. and {Parikh}, M.~M. and {Nair}, P.~H. and {Unther}, H.~M. and {Deil}, C. and {Woillez}, J. and {Conseil}, S. and {Kramer}, R. and {Turner}, J.~E.~H. and {Singer}, L. and {Fox}, R. and {Weaver}, B.~A. and {Zabalza}, V. and {Edwards}, Z.~I. and {Azalee Bostroem}, K. and {Burke}, D.~J. and {Casey}, A.~R. and {Crawford}, S.~M. and {Dencheva}, N. and {Ely}, J. and {Jenness}, T. and {Labrie}, K. and {Lim}, P.~L. and {Pierfederici}, F. and {Pontzen}, A. and {Ptak}, A. and {Refsdal}, B. and {Servillat}, M. and {Streicher}, O.},
Doi = {10.1051/0004-6361/201322068},
Eid = {A33},
Eprint = {1307.6212},
Journal = {\aap},
Month = oct,
Pages = {A33},
Primaryclass = {astro-ph.IM},
Title = {{Astropy: A community Python package for astronomy}},
Volume = 558,
Year = 2013}

@ARTICLE{astropy:2018,
       author = {{Astropy Collaboration} and {Price-Whelan}, A.~M. and
         {Sip{\H{o}}cz}, B.~M. and {G{\"u}nther}, H.~M. and {Lim}, P.~L. and
         {Crawford}, S.~M. and {Conseil}, S. and {Shupe}, D.~L. and
         {Craig}, M.~W. and {Dencheva}, N. and {Ginsburg}, A. and {Vand
        erPlas}, J.~T. and {Bradley}, L.~D. and {P{\'e}rez-Su{\'a}rez}, D. and
         {de Val-Borro}, M. and {Aldcroft}, T.~L. and {Cruz}, K.~L. and
         {Robitaille}, T.~P. and {Tollerud}, E.~J. and {Ardelean}, C. and
         {Babej}, T. and {Bach}, Y.~P. and {Bachetti}, M. and {Bakanov}, A.~V. and
         {Bamford}, S.~P. and {Barentsen}, G. and {Barmby}, P. and
         {Baumbach}, A. and {Berry}, K.~L. and {Biscani}, F. and {Boquien}, M. and
         {Bostroem}, K.~A. and {Bouma}, L.~G. and {Brammer}, G.~B. and
         {Bray}, E.~M. and {Breytenbach}, H. and {Buddelmeijer}, H. and
         {Burke}, D.~J. and {Calderone}, G. and {Cano Rodr{\'\i}guez}, J.~L. and
         {Cara}, M. and {Cardoso}, J.~V.~M. and {Cheedella}, S. and {Copin}, Y. and
         {Corrales}, L. and {Crichton}, D. and {D'Avella}, D. and {Deil}, C. and
         {Depagne}, {\'E}. and {Dietrich}, J.~P. and {Donath}, A. and
         {Droettboom}, M. and {Earl}, N. and {Erben}, T. and {Fabbro}, S. and
         {Ferreira}, L.~A. and {Finethy}, T. and {Fox}, R.~T. and
         {Garrison}, L.~H. and {Gibbons}, S.~L.~J. and {Goldstein}, D.~A. and
         {Gommers}, R. and {Greco}, J.~P. and {Greenfield}, P. and
         {Groener}, A.~M. and {Grollier}, F. and {Hagen}, A. and {Hirst}, P. and
         {Homeier}, D. and {Horton}, A.~J. and {Hosseinzadeh}, G. and {Hu}, L. and
         {Hunkeler}, J.~S. and {Ivezi{\'c}}, {\v{Z}}. and {Jain}, A. and
         {Jenness}, T. and {Kanarek}, G. and {Kendrew}, S. and {Kern}, N.~S. and
         {Kerzendorf}, W.~E. and {Khvalko}, A. and {King}, J. and {Kirkby}, D. and
         {Kulkarni}, A.~M. and {Kumar}, A. and {Lee}, A. and {Lenz}, D. and
         {Littlefair}, S.~P. and {Ma}, Z. and {Macleod}, D.~M. and
         {Mastropietro}, M. and {McCully}, C. and {Montagnac}, S. and
         {Morris}, B.~M. and {Mueller}, M. and {Mumford}, S.~J. and {Muna}, D. and
         {Murphy}, N.~A. and {Nelson}, S. and {Nguyen}, G.~H. and
         {Ninan}, J.~P. and {N{\"o}the}, M. and {Ogaz}, S. and {Oh}, S. and
         {Parejko}, J.~K. and {Parley}, N. and {Pascual}, S. and {Patil}, R. and
         {Patil}, A.~A. and {Plunkett}, A.~L. and {Prochaska}, J.~X. and
         {Rastogi}, T. and {Reddy Janga}, V. and {Sabater}, J. and
         {Sakurikar}, P. and {Seifert}, M. and {Sherbert}, L.~E. and
         {Sherwood-Taylor}, H. and {Shih}, A.~Y. and {Sick}, J. and
         {Silbiger}, M.~T. and {Singanamalla}, S. and {Singer}, L.~P. and
         {Sladen}, P.~H. and {Sooley}, K.~A. and {Sornarajah}, S. and
         {Streicher}, O. and {Teuben}, P. and {Thomas}, S.~W. and
         {Tremblay}, G.~R. and {Turner}, J.~E.~H. and {Terr{\'o}n}, V. and
         {van Kerkwijk}, M.~H. and {de la Vega}, A. and {Watkins}, L.~L. and
         {Weaver}, B.~A. and {Whitmore}, J.~B. and {Woillez}, J. and
         {Zabalza}, V. and {Astropy Contributors}},
        title = "{The Astropy Project: Building an Open-science Project and Status of the v2.0 Core Package}",
      journal = {\aj},
         year = 2018,
        month = sep,
       volume = {156},
       number = {3},
          eid = {123},
        pages = {123},
          doi = {10.3847/1538-3881/aabc4f},
archivePrefix = {arXiv},
       eprint = {1801.02634},
 primaryClass = {astro-ph.IM},
       adsurl = {https://ui.adsabs.harvard.edu/abs/2018AJ....156..123A}
}

@ARTICLE{astropy:2022,
       author = {{Astropy Collaboration} and {Price-Whelan}, Adrian M. and {Lim}, Pey Lian and {Earl}, Nicholas and {Starkman}, Nathaniel and {Bradley}, Larry and {Shupe}, David L. and {Patil}, Aarya A. and {Corrales}, Lia and {Brasseur}, C.~E. and {N{"o}the}, Maximilian and {Donath}, Axel and {Tollerud}, Erik and {Morris}, Brett M. and {Ginsburg}, Adam and {Vaher}, Eero and {Weaver}, Benjamin A. and {Tocknell}, James and {Jamieson}, William and {van Kerkwijk}, Marten H. and {Robitaille}, Thomas P. and {Merry}, Bruce and {Bachetti}, Matteo and {G{"u}nther}, H. Moritz and {Aldcroft}, Thomas L. and {Alvarado-Montes}, Jaime A. and {Archibald}, Anne M. and {B{'o}di}, Attila and {Bapat}, Shreyas and {Barentsen}, Geert and {Baz{'a}n}, Juanjo and {Biswas}, Manish and {Boquien}, M{'e}d{'e}ric and {Burke}, D.~J. and {Cara}, Daria and {Cara}, Mihai and {Conroy}, Kyle E. and {Conseil}, Simon and {Craig}, Matthew W. and {Cross}, Robert M. and {Cruz}, Kelle L. and {D'Eugenio}, Francesco and {Dencheva}, Nadia and {Devillepoix}, Hadrien A.~R. and {Dietrich}, J{"o}rg P. and {Eigenbrot}, Arthur Davis and {Erben}, Thomas and {Ferreira}, Leonardo and {Foreman-Mackey}, Daniel and {Fox}, Ryan and {Freij}, Nabil and {Garg}, Suyog and {Geda}, Robel and {Glattly}, Lauren and {Gondhalekar}, Yash and {Gordon}, Karl D. and {Grant}, David and {Greenfield}, Perry and {Groener}, Austen M. and {Guest}, Steve and {Gurovich}, Sebastian and {Handberg}, Rasmus and {Hart}, Akeem and {Hatfield-Dodds}, Zac and {Homeier}, Derek and {Hosseinzadeh}, Griffin and {Jenness}, Tim and {Jones}, Craig K. and {Joseph}, Prajwel and {Kalmbach}, J. Bryce and {Karamehmetoglu}, Emir and {Ka{l}uszy{'n}ski}, Miko{l}aj and {Kelley}, Michael S.~P. and {Kern}, Nicholas and {Kerzendorf}, Wolfgang E. and {Koch}, Eric W. and {Kulumani}, Shankar and {Lee}, Antony and {Ly}, Chun and {Ma}, Zhiyuan and {MacBride}, Conor and {Maljaars}, Jakob M. and {Muna}, Demitri and {Murphy}, N.~A. and {Norman}, Henrik and {O'Steen}, Richard and {Oman}, Kyle A. and {Pacifici}, Camilla and {Pascual}, Sergio and {Pascual-Granado}, J. and {Patil}, Rohit R. and {Perren}, Gabriel I. and {Pickering}, Timothy E. and {Rastogi}, Tanuj and {Roulston}, Benjamin R. and {Ryan}, Daniel F. and {Rykoff}, Eli S. and {Sabater}, Jose and {Sakurikar}, Parikshit and {Salgado}, Jes{'u}s and {Sanghi}, Aniket and {Saunders}, Nicholas and {Savchenko}, Volodymyr and {Schwardt}, Ludwig and {Seifert-Eckert}, Michael and {Shih}, Albert Y. and {Jain}, Anany Shrey and {Shukla}, Gyanendra and {Sick}, Jonathan and {Simpson}, Chris and {Singanamalla}, Sudheesh and {Singer}, Leo P. and {Singhal}, Jaladh and {Sinha}, Manodeep and {Sip{H{o}}cz}, Brigitta M. and {Spitler}, Lee R. and {Stansby}, David and {Streicher}, Ole and {{{S}}umak}, Jani and {Swinbank}, John D. and {Taranu}, Dan S. and {Tewary}, Nikita and {Tremblay}, Grant R. and {Val-Borro}, Miguel de and {Van Kooten}, Samuel J. and {Vasovi{'c}}, Zlatan and {Verma}, Shresth and {de Miranda Cardoso}, Jos{'e} Vin{'i}cius and {Williams}, Peter K.~G. and {Wilson}, Tom J. and {Winkel}, Benjamin and {Wood-Vasey}, W.~M. and {Xue}, Rui and {Yoachim}, Peter and {Zhang}, Chen and {Zonca}, Andrea and {Astropy Project Contributors}},
        title = "{The Astropy Project: Sustaining and Growing a Community-oriented Open-source Project and the Latest Major Release (v5.0) of the Core Package}",
      journal = {\apj},
         year = 2022,
        month = aug,
       volume = {935},
       number = {2},
          eid = {167},
        pages = {167},
          doi = {10.3847/1538-4357/ac7c74},
archivePrefix = {arXiv},
       eprint = {2206.14220},
 primaryClass = {astro-ph.IM},
       adsurl = {https://ui.adsabs.harvard.edu/abs/2022ApJ...935..167A}
}

@ARTICLE{2003MNRAS.340..160B,
       author = {{Benson}, A.~J. and {Hoyle}, Fiona and {Torres}, Fernando and {Vogeley}, Michael S.},
        title = "{Galaxy voids in cold dark matter universes}",
      journal = {\mnras},
         year = 2003,
        month = mar,
       volume = {340},
       number = {1},
        pages = {160-174},
          doi = {10.1046/j.1365-8711.2003.06281.x},
archivePrefix = {arXiv},
       eprint = {astro-ph/0208257},
 primaryClass = {astro-ph},
       adsurl = {https://ui.adsabs.harvard.edu/abs/2003MNRAS.340..160B}
}

@ARTICLE{2025A&A...698A.260B,
       author = {{Bidaran}, Bahar and {de Daniloff}, Simon and {P{\'e}rez}, Isabel and {Zurita}, Almudena and {Rom{\'a}n}, Javier and {Argudo-Fern{\'a}ndez}, Mar{\'\i}a and {Espada}, Daniel and {Ruiz-Lara}, Tom{\'a}s and {S{\'a}nchez-Menguiano}, Laura and {Garc{\'\i}a-Benito}, Rub{\'e}n and {Peletier}, Reynier F. and {Ferr{\'e}-Mateu}, Anna and {Duarte Puertas}, Salvador and {Verley}, Simon and {Falc{\'o}n-Barroso}, Jes{\'u}s and {Florido}, Estrella and {Torres-R{\'\i}os}, Gloria and {Lisenfeld}, Ute and {Rela{\~n}o}, M{\'o}nica and {Jim{\'e}nez}, Andoni},
        title = "{Rendezvous in CAVITY: Kinematics and gas properties of an isolated dwarf-dwarf merging pair in a cosmic void region}",
      journal = {\aap},
         year = 2025,
        month = jun,
       volume = {698},
          eid = {A260},
        pages = {A260},
          doi = {10.1051/0004-6361/202453556},
archivePrefix = {arXiv},
       eprint = {2504.15359},
 primaryClass = {astro-ph.GA},
       adsurl = {https://ui.adsabs.harvard.edu/abs/2025A&A...698A.260B}
}

@ARTICLE{2025A&A...693L..16B,
       author = {{Bidaran}, Bahar and {P{\'e}rez}, Isabel and {S{\'a}nchez-Menguiano}, Laura and {Argudo-Fern{\'a}ndez}, Mar{\'\i}a and {Ferr{\'e}-Mateu}, Anna and {Navarro}, Julio F. and {Peletier}, Reynier F. and {Ruiz-Lara}, Tom{\'a}s and {van de Ven}, Glenn and {Verley}, Simon and {Zurita}, Almudena and {Duarte Puertas}, Salvador and {Falc{\'o}n-Barroso}, Jes{\'u}s and {S{\'a}nchez-Bl{\'a}zquez}, Patricia and {Jim{\'e}nez}, Andoni},
        title = "{The puzzle of isolated and quenched dwarf galaxies in cosmic voids}",
      journal = {\aap},
         year = 2025,
        month = jan,
       volume = {693},
          eid = {L16},
        pages = {L16},
          doi = {10.1051/0004-6361/202452688},
archivePrefix = {arXiv},
       eprint = {2501.02910},
 primaryClass = {astro-ph.GA},
       adsurl = {https://ui.adsabs.harvard.edu/abs/2025A&A...693L..16B}
}

@ARTICLE{2007AJ....133..734B,
   author = {{Blanton}, M.~R. and {Roweis}, S.},
    title = "{K-Corrections and Filter Transformations in the Ultraviolet, Optical, and Near-Infrared}",
  journal = {\aj},
   eprint = {astro-ph/0606170},
     year = 2007,
    month = feb,
   volume = 133,
    pages = {734-754},
      doi = {10.1086/510127},
   adsurl = {http://adsabs.harvard.edu/abs/2007AJ....133..734B}
}

@ARTICLE{1996Natur.380..603B,
       author = {{Bond}, J. Richard and {Kofman}, Lev and {Pogosyan}, Dmitry},
        title = "{How filaments of galaxies are woven into the cosmic web}",
      journal = {\nat},
         year = 1996,
        month = apr,
       volume = {380},
       number = {6575},
        pages = {603-606},
          doi = {10.1038/380603a0},
archivePrefix = {arXiv},
       eprint = {astro-ph/9512141},
 primaryClass = {astro-ph},
       adsurl = {https://ui.adsabs.harvard.edu/abs/1996Natur.380..603B}
}

@ARTICLE{2014MNRAS.441.2923C,
       author = {{Cautun}, Marius and {van de Weygaert}, Rien and {Jones}, Bernard J.~T. and {Frenk}, Carlos S.},
        title = "{Evolution of the cosmic web}",
      journal = {\mnras},
         year = 2014,
        month = jul,
       volume = {441},
       number = {4},
        pages = {2923-2973},
          doi = {10.1093/mnras/stu768},
archivePrefix = {arXiv},
       eprint = {1401.7866},
 primaryClass = {astro-ph.CO},
       adsurl = {https://ui.adsabs.harvard.edu/abs/2014MNRAS.441.2923C}
}

@ARTICLE{2025A&A...700A.196C,
       author = {{Ceccarelli}, Maria Laura and {Alonso}, Sol and {Garcia Lambas}, Diego},
        title = "{Galaxy pairs in cosmic voids}",
      journal = {\aap},
         year = 2025,
        month = aug,
       volume = {700},
          eid = {A196},
        pages = {A196},
          doi = {10.1051/0004-6361/202453369},
archivePrefix = {arXiv},
       eprint = {2412.01697},
 primaryClass = {astro-ph.GA},
       adsurl = {https://ui.adsabs.harvard.edu/abs/2025A&A...700A.196C}
}

@ARTICLE{2008MNRAS.390L...9C,
       author = {{Ceccarelli}, L. and {Padilla}, N. and {Lambas}, D.~G.},
        title = "{Large-scale modulation of star formation in void walls}",
      journal = {\mnras},
         year = 2008,
        month = oct,
       volume = {390},
       number = {3},
        pages = {L9-L13},
          doi = {10.1111/j.1745-3933.2008.00520.x},
archivePrefix = {arXiv},
       eprint = {0805.0790},
 primaryClass = {astro-ph},
       adsurl = {https://ui.adsabs.harvard.edu/abs/2008MNRAS.390L...9C}
}

@ARTICLE{2013MNRAS.434.1435C,
       author = {{Ceccarelli}, L. and {Paz}, D. and {Lares}, M. and {Padilla}, N. and {Lambas}, D. Garc{\'\i}a},
        title = "{Clues on void evolution - I. Large-scale galaxy distributions around voids}",
      journal = {\mnras},
         year = 2013,
        month = sep,
       volume = {434},
       number = {2},
        pages = {1435-1442},
          doi = {10.1093/mnras/stt1097},
archivePrefix = {arXiv},
       eprint = {1306.5798},
 primaryClass = {astro-ph.CO},
       adsurl = {https://ui.adsabs.harvard.edu/abs/2013MNRAS.434.1435C}
}

@ARTICLE{2024A&A...687A..98C,
       author = {{Conrado}, Ana M. and {Gonz{\'a}lez Delgado}, Rosa M. and {Garc{\'\i}a-Benito}, Rub{\'e}n and {P{\'e}rez}, Isabel and {Verley}, Simon and {Ruiz-Lara}, Tom{\'a}s and {S{\'a}nchez-Menguiano}, Laura and {Duarte Puertas}, Salvador and {Jim{\'e}nez}, Andoni and {Dom{\'\i}nguez-G{\'o}mez}, Jes{\'u}s and {Espada}, Daniel and {Argudo-Fern{\'a}ndez}, Mar{\'\i}a and {Alc{\'a}zar-Laynez}, Manuel and {Bl{\'a}zquez-Calero}, Guillermo and {Bidaran}, Bahar and {Zurita}, Almudena and {Peletier}, Reynier and {Torres-R{\'\i}os}, Gloria and {Florido}, Estrella and {Rodr{\'\i}guez Mart{\'\i}nez}, M{\'o}nica and {del Moral-Castro}, Ignacio and {van de Weygaert}, Rien and {Falc{\'o}n-Barroso}, Jes{\'u}s and {Lugo-Aranda}, Alejandra Z. and {S{\'a}nchez}, Sebasti{\'a}n F. and {van der Hulst}, Thijs and {Courtois}, H{\'e}l{\`e}ne M. and {Ferr{\'e}-Mateu}, Anna and {S{\'a}nchez-Bl{\'a}zquez}, Patricia and {Rom{\'a}n}, Javier and {Aceituno}, Jes{\'u}s},
        title = "{The CAVITY project: The spatially resolved stellar population properties of galaxies in voids}",
      journal = {\aap},
         year = 2024,
        month = jul,
       volume = {687},
          eid = {A98},
        pages = {A98},
          doi = {10.1051/0004-6361/202449414},
archivePrefix = {arXiv},
       eprint = {2404.10823},
 primaryClass = {astro-ph.GA},
       adsurl = {https://ui.adsabs.harvard.edu/abs/2024A&A...687A..98C}
}

@ARTICLE{2005MNRAS.356.1155C,
       author = {{Croton}, Darren J. and {Farrar}, Glennys R. and {Norberg}, Peder and {Colless}, Matthew and {Peacock}, John A. and {Baldry}, I.~K. and {Baugh}, C.~M. and {Bland-Hawthorn}, J. and {Bridges}, T. and {Cannon}, R. and {Cole}, S. and {Collins}, C. and {Couch}, W. and {Dalton}, G. and {De Propris}, R. and {Driver}, S.~P. and {Efstathiou}, G. and {Ellis}, R.~S. and {Frenk}, C.~S. and {Glazebrook}, K. and {Jackson}, C. and {Lahav}, O. and {Lewis}, I. and {Lumsden}, S. and {Maddox}, S. and {Madgwick}, D. and {Peterson}, B.~A. and {Sutherland}, W. and {Taylor}, K.},
        title = "{The 2dF Galaxy Redshift Survey: luminosity functions by density environment and galaxy type}",
      journal = {\mnras},
         year = 2005,
        month = jan,
       volume = {356},
       number = {3},
        pages = {1155-1167},
          doi = {10.1111/j.1365-2966.2004.08546.x},
archivePrefix = {arXiv},
       eprint = {astro-ph/0407537},
 primaryClass = {astro-ph},
       adsurl = {https://ui.adsabs.harvard.edu/abs/2005MNRAS.356.1155C}
}

@ARTICLE{2024ApJ...962...58C,
       author = {{Curtis}, Olivia and {McDonough}, Bryanne and {Brainerd}, Tereasa G.},
        title = "{Properties of Voids and Void Galaxies in the TNG300 Simulation}",
      journal = {\apj},
         year = 2024,
        month = feb,
       volume = {962},
       number = {1},
          eid = {58},
        pages = {58},
          doi = {10.3847/1538-4357/ad18b4},
archivePrefix = {arXiv},
       eprint = {2401.02322},
 primaryClass = {astro-ph.GA},
       adsurl = {https://ui.adsabs.harvard.edu/abs/2024ApJ...962...58C}
}

@ARTICLE{2007A&A...474..783D,
       author = {{Deng}, Xin-Fa and {He}, Ji-Zhou and {Jiang}, Peng and {Luo}, Cheng-Hong and {Wu}, Ping},
        title = "{The main galaxy groups from the SDSS data release 5}",
      journal = {\aap},
         year = 2007,
        month = nov,
       volume = {474},
       number = {3},
        pages = {783-791},
          doi = {10.1051/0004-6361:20066407},
       adsurl = {https://ui.adsabs.harvard.edu/abs/2007A&A...474..783D}
}

@ARTICLE{1993AJ....105.2035D,
       author = {{Diaferio}, Antonaldo and {Ramella}, Massimo and {Geller}, Margaret J. and {Ferrari}, Attilio},
        title = "{Are Groups of Galaxies Virialized Systems?}",
      journal = {\aj},
         year = 1993,
        month = jun,
       volume = {105},
        pages = {2035},
          doi = {10.1086/116581},
       adsurl = {https://ui.adsabs.harvard.edu/abs/1993AJ....105.2035D}
}

@ARTICLE{2023Natur.619..269D,
       author = {{Dom{\'\i}nguez-G{\'o}mez}, Jes{\'u}s and {P{\'e}rez}, Isabel and {Ruiz-Lara}, Tom{\'a}s and {Peletier}, Reynier F. and {S{\'a}nchez-Bl{\'a}zquez}, Patricia and {Lisenfeld}, Ute and {Falc{\'o}n-Barroso}, Jes{\'u}s and {Alc{\'a}zar-Laynez}, Manuel and {Argudo-Fern{\'a}ndez}, Mar{\'\i}a and {Bl{\'a}zquez-Calero}, Guillermo and {Courtois}, H{\'e}l{\`e}ne and {Duarte Puertas}, Salvador and {Espada}, Daniel and {Florido}, Estrella and {Garc{\'\i}a-Benito}, Rub{\'e}n and {Jim{\'e}nez}, Andoni and {Kreckel}, Kathryn and {Rela{\~n}o}, M{\'o}nica and {S{\'a}nchez-Menguiano}, Laura and {van der Hulst}, Thijs and {van de Weygaert}, Rien and {Verley}, Simon and {Zurita}, Almudena},
        title = "{Galaxies in voids assemble their stars slowly}",
      journal = {\nat},
         year = 2023,
        month = jul,
       volume = {619},
       number = {7969},
        pages = {269-271},
          doi = {10.1038/s41586-023-06109-1},
archivePrefix = {arXiv},
       eprint = {2306.16818},
 primaryClass = {astro-ph.GA},
       adsurl = {https://ui.adsabs.harvard.edu/abs/2023Natur.619..269D}
}

@ARTICLE{2023A&A...680A.111D,
       author = {{Dom{\'\i}nguez-G{\'o}mez}, Jes{\'u}s and {P{\'e}rez}, Isabel and {Ruiz-Lara}, Tom{\'a}s and {Peletier}, Reynier F. and {S{\'a}nchez-Bl{\'a}zquez}, Patricia and {Lisenfeld}, Ute and {Bidaran}, Bahar and {Falc{\'o}n-Barroso}, Jes{\'u}s and {Alc{\'a}zar-Laynez}, Manuel and {Argudo-Fern{\'a}ndez}, Mar{\'\i}a and {Bl{\'a}zquez-Calero}, Guillermo and {Courtois}, H{\'e}l{\`e}ne and {Duarte Puertas}, Salvador and {Espada}, Daniel and {Florido}, Estrella and {Garc{\'\i}a-Benito}, Rub{\'e}n and {Jim{\'e}nez}, Andoni and {Kreckel}, Kathryn and {Rela{\~n}o}, M{\'o}nica and {S{\'a}nchez-Menguiano}, Laura and {van der Hulst}, Thijs and {van de Weygaert}, Rien and {Verley}, Simon and {Zurita}, Almudena},
        title = "{Stellar mass-metallicity relation throughout the large-scale structure of the Universe: CAVITY mother sample}",
      journal = {\aap},
         year = 2023,
        month = dec,
       volume = {680},
          eid = {A111},
        pages = {A111},
          doi = {10.1051/0004-6361/202346884},
archivePrefix = {arXiv},
       eprint = {2310.11412},
 primaryClass = {astro-ph.GA},
       adsurl = {https://ui.adsabs.harvard.edu/abs/2023A&A...680A.111D}
}

@ARTICLE{2009A&G....50e..12D,
       author = {{Driver}, Simon P. and {Norberg}, Peder and {Baldry}, Ivan K. and {Bamford}, Steven P. and {Hopkins}, Andrew M. and {Liske}, Jochen and {Loveday}, Jon and {Peacock}, John A. and {Hill}, D.~T. and {Kelvin}, L.~S. and {Robotham}, A.~S.~G. and {Cross}, N.~J.~G. and {Parkinson}, H.~R. and {Prescott}, M. and {Conselice}, C.~J. and {Dunne}, L. and {Brough}, S. and {Jones}, H. and {Sharp}, R.~G. and {van Kampen}, E. and {Oliver}, S. and {Roseboom}, I.~G. and {Bland-Hawthorn}, J. and {Croom}, S.~M. and {Ellis}, S. and {Cameron}, E. and {Cole}, S. and {Frenk}, C.~S. and {Couch}, W.~J. and {Graham}, A.~W. and {Proctor}, R. and {De Propris}, R. and {Doyle}, I.~F. and {Edmondson}, E.~M. and {Nichol}, R.~C. and {Thomas}, D. and {Eales}, S.~A. and {Jarvis}, M.~J. and {Kuijken}, K. and {Lahav}, O. and {Madore}, B.~F. and {Seibert}, M. and {Meyer}, M.~J. and {Staveley-Smith}, L. and {Phillipps}, S. and {Popescu}, C.~C. and {Sansom}, A.~E. and {Sutherland}, W.~J. and {Tuffs}, R.~J. and {Warren}, S.~J.},
        title = "{GAMA: towards a physical understanding of galaxy formation}",
      journal = {Astronomy and Geophysics},
         year = 2009,
        month = oct,
       volume = {50},
       number = {5},
        pages = {5.12-5.19},
          doi = {10.1111/j.1468-4004.2009.50512.x},
archivePrefix = {arXiv},
       eprint = {0910.5123},
 primaryClass = {astro-ph.CO},
       adsurl = {https://ui.adsabs.harvard.edu/abs/2009A&G....50e..12D}
}

@ARTICLE{2011MNRAS.413..971D,
       author = {{Driver}, S.~P. and {Hill}, D.~T. and {Kelvin}, L.~S. and {Robotham}, A.~S.~G. and {Liske}, J. and {Norberg}, P. and {Baldry}, I.~K. and {Bamford}, S.~P. and {Hopkins}, A.~M. and {Loveday}, J. and {Peacock}, J.~A. and {Andrae}, E. and {Bland-Hawthorn}, J. and {Brough}, S. and {Brown}, M.~J.~I. and {Cameron}, E. and {Ching}, J.~H.~Y. and {Colless}, M. and {Conselice}, C.~J. and {Croom}, S.~M. and {Cross}, N.~J.~G. and {de Propris}, R. and {Dye}, S. and {Drinkwater}, M.~J. and {Ellis}, S. and {Graham}, Alister W. and {Grootes}, M.~W. and {Gunawardhana}, M. and {Jones}, D.~H. and {van Kampen}, E. and {Maraston}, C. and {Nichol}, R.~C. and {Parkinson}, H.~R. and {Phillipps}, S. and {Pimbblet}, K. and {Popescu}, C.~C. and {Prescott}, M. and {Roseboom}, I.~G. and {Sadler}, E.~M. and {Sansom}, A.~E. and {Sharp}, R.~G. and {Smith}, D.~J.~B. and {Taylor}, E. and {Thomas}, D. and {Tuffs}, R.~J. and {Wijesinghe}, D. and {Dunne}, L. and {Frenk}, C.~S. and {Jarvis}, M.~J. and {Madore}, B.~F. and {Meyer}, M.~J. and {Seibert}, M. and {Staveley-Smith}, L. and {Sutherland}, W.~J. and {Warren}, S.~J.},
        title = "{Galaxy and Mass Assembly (GAMA): survey diagnostics and core data release}",
      journal = {\mnras},
         year = 2011,
        month = may,
       volume = {413},
       number = {2},
        pages = {971-995},
          doi = {10.1111/j.1365-2966.2010.18188.x},
archivePrefix = {arXiv},
       eprint = {1009.0614},
 primaryClass = {astro-ph.CO},
       adsurl = {https://ui.adsabs.harvard.edu/abs/2011MNRAS.413..971D}
}

@ARTICLE{2024A&A...681A..91E,
       author = {{Einasto}, Maret and {Einasto}, Jaan and {Tenjes}, Peeter and {Korhonen}, Suvi and {Kipper}, Rain and {Tempel}, Elmo and {Liivam{\"a}gi}, Lauri Juhan and {Hein{\"a}m{\"a}ki}, Pekka},
        title = "{Galaxy groups and clusters and their brightest galaxies within the cosmic web}",
      journal = {\aap},
         year = 2024,
        month = jan,
       volume = {681},
          eid = {A91},
        pages = {A91},
          doi = {10.1051/0004-6361/202347504},
archivePrefix = {arXiv},
       eprint = {2311.01868},
 primaryClass = {astro-ph.CO},
       adsurl = {https://ui.adsabs.harvard.edu/abs/2024A&A...681A..91E}
}

@ARTICLE{2003A&A...401..851E,
       author = {{Einasto}, M. and {Einasto}, J. and {M{\"u}ller}, V. and {Hein{\"a}m{\"a}ki}, P. and {Tucker}, D.~L.},
        title = "{Environmental enhancement of loose groups  around rich clusters of galaxies}",
      journal = {\aap},
         year = 2003,
        month = apr,
       volume = {401},
        pages = {851-862},
          doi = {10.1051/0004-6361:20021727},
archivePrefix = {arXiv},
       eprint = {astro-ph/0211590},
 primaryClass = {astro-ph},
       adsurl = {https://ui.adsabs.harvard.edu/abs/2003A&A...401..851E}
}

@ARTICLE{2011AJ....142...72E,
       author = {{Eisenstein}, Daniel J. and {Weinberg}, David H. and {Agol}, Eric and {Aihara}, Hiroaki and {Allende Prieto}, Carlos and {Anderson}, Scott F. and {Arns}, James A. and {Aubourg}, {\'E}ric and {Bailey}, Stephen and {Balbinot}, Eduardo and {Barkhouser}, Robert and {Beers}, Timothy C. and {Berlind}, Andreas A. and {Bickerton}, Steven J. and {Bizyaev}, Dmitry and {Blanton}, Michael R. and {Bochanski}, John J. and {Bolton}, Adam S. and {Bosman}, Casey T. and {Bovy}, Jo and {Brandt}, W.~N. and {Breslauer}, Ben and {Brewington}, Howard J. and {Brinkmann}, J. and {Brown}, Peter J. and {Brownstein}, Joel R. and {Burger}, Dan and {Busca}, Nicolas G. and {Campbell}, Heather and {Cargile}, Phillip A. and {Carithers}, William C. and {Carlberg}, Joleen K. and {Carr}, Michael A. and {Chang}, Liang and {Chen}, Yanmei and {Chiappini}, Cristina and {Comparat}, Johan and {Connolly}, Natalia and {Cortes}, Marina and {Croft}, Rupert A.~C. and {Cunha}, Katia and {da Costa}, Luiz N. and {Davenport}, James R.~A. and {Dawson}, Kyle and {De Lee}, Nathan and {Porto de Mello}, Gustavo F. and {de Simoni}, Fernando and {Dean}, Janice and {Dhital}, Saurav and {Ealet}, Anne and {Ebelke}, Garrett L. and {Edmondson}, Edward M. and {Eiting}, Jacob M. and {Escoffier}, Stephanie and {Esposito}, Massimiliano and {Evans}, Michael L. and {Fan}, Xiaohui and {Femen{\'\i}a Castell{\'a}}, Bruno and {Dutra Ferreira}, Leticia and {Fitzgerald}, Greg and {Fleming}, Scott W. and {Font-Ribera}, Andreu and {Ford}, Eric B. and {Frinchaboy}, Peter M. and {Garc{\'\i}a P{\'e}rez}, Ana Elia and {Gaudi}, B. Scott and {Ge}, Jian and {Ghezzi}, Luan and {Gillespie}, Bruce A. and {Gilmore}, G. and {Girardi}, L{\'e}o and {Gott}, J. Richard and {Gould}, Andrew and {Grebel}, Eva K. and {Gunn}, James E. and {Hamilton}, Jean-Christophe and {Harding}, Paul and {Harris}, David W. and {Hawley}, Suzanne L. and {Hearty}, Frederick R. and {Hennawi}, Joseph F. and {Gonz{\'a}lez Hern{\'a}ndez}, Jonay I. and {Ho}, Shirley and {Hogg}, David W. and {Holtzman}, Jon A. and {Honscheid}, Klaus and {Inada}, Naohisa and {Ivans}, Inese I. and {Jiang}, Linhua and {Jiang}, Peng and {Johnson}, Jennifer A. and {Jordan}, Cathy and {Jordan}, Wendell P. and {Kauffmann}, Guinevere and {Kazin}, Eyal and {Kirkby}, David and {Klaene}, Mark A. and {Knapp}, G.~R. and {Kneib}, Jean-Paul and {Kochanek}, C.~S. and {Koesterke}, Lars and {Kollmeier}, Juna A. and {Kron}, Richard G. and {Lampeitl}, Hubert and {Lang}, Dustin and {Lawler}, James E. and {Le Goff}, Jean-Marc and {Lee}, Brian L. and {Lee}, Young Sun and {Leisenring}, Jarron M. and {Lin}, Yen-Ting and {Liu}, Jian and {Long}, Daniel C. and {Loomis}, Craig P. and {Lucatello}, Sara and {Lundgren}, Britt and {Lupton}, Robert H. and {Ma}, Bo and {Ma}, Zhibo and {MacDonald}, Nicholas and {Mack}, Claude and {Mahadevan}, Suvrath and {Maia}, Marcio A.~G. and {Majewski}, Steven R. and {Makler}, Martin and {Malanushenko}, Elena and {Malanushenko}, Viktor and {Mandelbaum}, Rachel and {Maraston}, Claudia and {Margala}, Daniel and {Maseman}, Paul and {Masters}, Karen L. and {McBride}, Cameron K. and {McDonald}, Patrick and {McGreer}, Ian D. and {McMahon}, Richard G. and {Mena Requejo}, Olga and {M{\'e}nard}, Brice and {Miralda-Escud{\'e}}, Jordi and {Morrison}, Heather L. and {Mullally}, Fergal and {Muna}, Demitri and {Murayama}, Hitoshi and {Myers}, Adam D. and {Naugle}, Tracy and {Neto}, Angelo Fausti and {Nguyen}, Duy Cuong and {Nichol}, Robert C. and {Nidever}, David L. and {O'Connell}, Robert W. and {Ogando}, Ricardo L.~C. and {Olmstead}, Matthew D. and {Oravetz}, Daniel J. and {Padmanabhan}, Nikhil and {Paegert}, Martin and {Palanque-Delabrouille}, Nathalie and {Pan}, Kaike and {Pandey}, Parul and {Parejko}, John K. and {P{\^a}ris}, Isabelle and {Pellegrini}, Paulo and {Pepper}, Joshua and {Percival}, Will J. and {Petitjean}, Patrick and {Pfaffenberger}, Robert and {Pforr}, Janine and {Phleps}, Stefanie and {Pichon}, Christophe and {Pieri}, Matthew M. and {Prada}, Francisco and {Price-Whelan}, Adrian M. and {Raddick}, M. Jordan and {Ramos}, Beatriz H.~F. and {Reid}, I. Neill and {Reyle}, Celine and {Rich}, James and {Richards}, Gordon T. and {Rieke}, George H. and {Rieke}, Marcia J. and {Rix}, Hans-Walter and {Robin}, Annie C. and {Rocha-Pinto}, Helio J. and {Rockosi}, Constance M. and {Roe}, Natalie A. and {Rollinde}, Emmanuel and {Ross}, Ashley J. and {Ross}, Nicholas P. and {Rossetto}, Bruno and {S{\'a}nchez}, Ariel G. and {Santiago}, Basilio and {Sayres}, Conor and {Schiavon}, Ricardo and {Schlegel}, David J. and {Schlesinger}, Katharine J. and {Schmidt}, Sarah J. and {Schneider}, Donald P. and {Sellgren}, Kris and {Shelden}, Alaina and {Sheldon}, Erin and {Shetrone}, Matthew},
        title = "{SDSS-III: Massive Spectroscopic Surveys of the Distant Universe, the Milky Way, and Extra-Solar Planetary Systems}",
      journal = {\aj},
         year = 2011,
        month = sep,
       volume = {142},
       number = {3},
          eid = {72},
        pages = {72},
          doi = {10.1088/0004-6256/142/3/72},
archivePrefix = {arXiv},
       eprint = {1101.1529},
 primaryClass = {astro-ph.IM},
       adsurl = {https://ui.adsabs.harvard.edu/abs/2011AJ....142...72E}
}

@ARTICLE{1997MNRAS.287..790E,
       author = {{El-Ad}, H. and {Piran}, T. and {Dacosta}, L.~N.},
        title = "{A catalogue of the voids in the IRAS 1.2-Jy survey}",
      journal = {\mnras},
         year = 1997,
        month = jun,
       volume = {287},
       number = {4},
        pages = {790-798},
          doi = {10.1093/mnras/287.4.790},
archivePrefix = {arXiv},
       eprint = {astro-ph/9608022},
 primaryClass = {astro-ph},
       adsurl = {https://ui.adsabs.harvard.edu/abs/1997MNRAS.287..790E}
}

@ARTICLE{2015MNRAS.451..660E,
       author = {{Etherington}, James and {Thomas}, Daniel},
        title = "{Measuring galaxy environments in large-scale photometric surveys}",
      journal = {\mnras},
         year = 2015,
        month = jul,
       volume = {451},
       number = {1},
        pages = {660-679},
          doi = {10.1093/mnras/stv999},
archivePrefix = {arXiv},
       eprint = {1505.01171},
 primaryClass = {astro-ph.GA},
       adsurl = {https://ui.adsabs.harvard.edu/abs/2015MNRAS.451..660E}
}

@ARTICLE{2016RAA....16...72F,
       author = {{Feng}, Shuai and {Shao}, Zheng-Yi and {Shen}, Shi-Yin and {Argudo-Fern{\'a}ndez}, Maria and {Wu}, Hong and {Lam}, Man-I. and {Yang}, Ming and {Yuan}, Fang-Ting},
        title = "{An isolated compact galaxy triplet}",
      journal = {Research in Astronomy and Astrophysics},
         year = 2016,
        month = may,
       volume = {16},
       number = {5},
          eid = {72},
        pages = {72},
          doi = {10.1088/1674-4527/16/5/072},
archivePrefix = {arXiv},
       eprint = {1512.02439},
 primaryClass = {astro-ph.GA},
       adsurl = {https://ui.adsabs.harvard.edu/abs/2016RAA....16...72F}
}

@ARTICLE{2021ApJ...906...97F,
       author = {{Florez}, Jonathan and {Berlind}, Andreas A. and {Kannappan}, Sheila J. and {Stark}, David V. and {Eckert}, Kathleen D. and {Calderon}, Victor F. and {Moffett}, Amanda J. and {Campbell}, Duncan and {Sinha}, Manodeep},
        title = "{Void Galaxies Follow a Distinct Evolutionary Path in the Environmental COntext Catalog}",
      journal = {\apj},
         year = 2021,
        month = jan,
       volume = {906},
       number = {2},
          eid = {97},
        pages = {97},
          doi = {10.3847/1538-4357/abca9f},
archivePrefix = {arXiv},
       eprint = {2011.08276},
 primaryClass = {astro-ph.GA},
       adsurl = {https://ui.adsabs.harvard.edu/abs/2021ApJ...906...97F}
}

@ARTICLE{2025arXiv250615345G,
       author = {{Gal{\'a}rraga-Espinosa}, Daniela and {Kauffmann}, Guinevere and {Bonoli}, Silvia and {Lucie-Smith}, Luisa and {Gonz{\'a}lez Delgado}, Rosa M. and {Tempel}, Elmo and {Abramo}, Raul and {Gurung-L{\'o}pez}, Siddharta and {Marra}, Valerio and {Alcaniz}, Jailson and {Benitez}, Narciso and {Carneiro}, Saulo and {Cenarro}, Javier and {Crist{\'o}bal-Hornillos}, David and {Dupke}, Renato and {Ederoclite}, Alessandro and {Hern{\'a}n-Caballero}, Antonio and {Hern{\'a}ndez-Monteagudo}, Carlos and {L{\'o}pez-Sanjuan}, Carlos and {Mar{\'\i}n-Franch}, Antonio and {Mendes de Oliveira}, Claudia and {Moles}, Mariano and {Sodr{\'e}}, Jr, Laerte and {Taylor}, Keith and {Varela}, Jes{\'u}s and {V{\'a}zquez Rami{\'o}}, Hector},
        title = "{Unveiling the small-scale web around galaxies with miniJPAS and DESI}",
      journal = {arXiv e-prints},
         year = 2025,
        month = jun,
          eid = {arXiv:2506.15345},
        pages = {arXiv:2506.15345},
          doi = {10.48550/arXiv.2506.15345},
archivePrefix = {arXiv},
       eprint = {2506.15345},
 primaryClass = {astro-ph.CO},
       adsurl = {https://ui.adsabs.harvard.edu/abs/2025arXiv250615345G}
}

@ARTICLE{2023A&A...671A.160G,
       author = {{Gal{\'a}rraga-Espinosa}, Daniela and {Garaldi}, Enrico and {Kauffmann}, Guinevere},
        title = "{Flows around galaxies. I. The dependence of galaxy connectivity on cosmic environments and effects on the star formation rate}",
      journal = {\aap},
         year = 2023,
        month = mar,
       volume = {671},
          eid = {A160},
        pages = {A160},
          doi = {10.1051/0004-6361/202244935},
archivePrefix = {arXiv},
       eprint = {2209.05495},
 primaryClass = {astro-ph.GA},
       adsurl = {https://ui.adsabs.harvard.edu/abs/2023A&A...671A.160G}
}

@ARTICLE{2011ApJ...728...74G,
       author = {{Galaz}, Gaspar and {Herrera-Camus}, Rodrigo and {Garcia-Lambas}, Diego and {Padilla}, Nelson},
        title = "{Low Surface Brightness Galaxies in the SDSS: The Link Between Environment, Star-forming Properties, and Active Galactic Nuclei}",
      journal = {\apj},
         year = 2011,
        month = feb,
       volume = {728},
       number = {2},
          eid = {74},
        pages = {74},
          doi = {10.1088/0004-637X/728/2/74},
archivePrefix = {arXiv},
       eprint = {1007.4014},
 primaryClass = {astro-ph.CO},
       adsurl = {https://ui.adsabs.harvard.edu/abs/2011ApJ...728...74G}
}

@ARTICLE{2024A&A...691A.161G,
       author = {{Garc{\'\i}a-Benito}, Rub{\'e}n and {Jim{\'e}nez}, Andoni and {S{\'a}nchez-Menguiano}, Laura and {Ruiz-Lara}, Tom{\'a}s and {Duarte Puertas}, Salvador and {Dom{\'\i}nguez-G{\'o}mez}, Jes{\'u}s and {Bidaran}, Bahar and {Torres-R{\'\i}os}, Gloria and {Argudo-Fern{\'a}ndez}, Mar{\'\i}a and {Espada}, Daniel and {P{\'e}rez}, Isabel and {Verley}, Simon and {Conrado}, Ana M. and {Florido}, Estrella and {Rodr{\'\i}guez}, M{\'o}nica I. and {Zurita}, Almudena and {Alc{\'a}zar-Laynez}, Manuel and {De Daniloff}, Simon B. and {Lisenfeld}, Ute and {van de Weygaert}, Rien and {Courtois}, H{\'e}l{\`e}ne M. and {Falc{\'o}n-Barroso}, Jes{\'u}s and {Ferr{\'e}-Mateu}, Anna and {Galbany}, Llu{\'\i}s and {Gonz{\'a}lez Delgado}, Rosa M. and {del Moral-Castro}, Ignacio and {Peletier}, Reynier F. and {Rom{\'a}n}, Javier and {S{\'a}nchez}, Sebasti{\'a}n F. and {S{\'a}nchez-Alarc{\'o}n}, Pablo M. and {S{\'a}nchez-Bl{\'a}zquez}, Patricia and {Villalba-Gonz{\'a}lez}, Pedro and {Azzaro}, Marco and {Blazek}, Mart{\'\i}n and {Fern{\'a}ndez}, Alba and {Gallego}, Julia and {G{\'o}ngora}, Samuel and {Guijarro}, Ana and {de Guindos}, Enrique and {Hermelo}, Israel and {Hern{\'a}ndez}, Ricardo and {de Juan}, Enrique and {Vico Linares}, Jos{\'e} Ignacio},
        title = "{CAVITY: Calar Alto Void Integral-field Treasury surveY: I. First public data release}",
      journal = {\aap},
         year = 2024,
        month = nov,
       volume = {691},
          eid = {A161},
        pages = {A161},
          doi = {10.1051/0004-6361/202451400},
archivePrefix = {arXiv},
       eprint = {2410.08265},
 primaryClass = {astro-ph.GA},
       adsurl = {https://ui.adsabs.harvard.edu/abs/2024A&A...691A.161G}
}

@ARTICLE{1999AJ....118.2561G,
       author = {{Grogin}, Norman A. and {Geller}, Margaret J.},
        title = "{An Imaging and Spectroscopic Survey of Galaxies within Prominent Nearby Voids. I. The Sample and Luminosity Distribution}",
      journal = {\aj},
         year = 1999,
        month = dec,
       volume = {118},
       number = {6},
        pages = {2561-2580},
          doi = {10.1086/301126},
archivePrefix = {arXiv},
       eprint = {astro-ph/9910073},
 primaryClass = {astro-ph},
       adsurl = {https://ui.adsabs.harvard.edu/abs/1999AJ....118.2561G}
}

@ARTICLE{2000AJ....119...32G,
       author = {{Grogin}, Norman A. and {Geller}, Margaret J.},
        title = "{An Imaging and Spectroscopic Survey of Galaxies within Prominent Nearby Voids. II. Morphologies, Star Formation, and Faint Companions}",
      journal = {\aj},
         year = 2000,
        month = jan,
       volume = {119},
       number = {1},
        pages = {32-43},
          doi = {10.1086/301179},
archivePrefix = {arXiv},
       eprint = {astro-ph/9910096},
 primaryClass = {astro-ph},
       adsurl = {https://ui.adsabs.harvard.edu/abs/2000AJ....119...32G}
}

@ARTICLE{2020MNRAS.493..899H,
       author = {{Habouzit}, M{\'e}lanie and {Pisani}, Alice and {Goulding}, Andy and {Dubois}, Yohan and {Somerville}, Rachel S. and {Greene}, Jenny E.},
        title = "{Properties of simulated galaxies and supermassive black holes in cosmic voids}",
      journal = {\mnras},
         year = 2020,
        month = mar,
       volume = {493},
       number = {1},
        pages = {899-921},
          doi = {10.1093/mnras/staa219},
archivePrefix = {arXiv},
       eprint = {1912.06662},
 primaryClass = {astro-ph.GA},
       adsurl = {https://ui.adsabs.harvard.edu/abs/2020MNRAS.493..899H}
}

@ARTICLE{2007MNRAS.381...41H,
       author = {{Hahn}, Oliver and {Carollo}, C. Marcella and {Porciani}, Cristiano and {Dekel}, Avishai},
        title = "{The evolution of dark matter halo properties in clusters, filaments, sheets and voids}",
      journal = {\mnras},
         year = 2007,
        month = oct,
       volume = {381},
       number = {1},
        pages = {41-51},
          doi = {10.1111/j.1365-2966.2007.12249.x},
archivePrefix = {arXiv},
       eprint = {0704.2595},
 primaryClass = {astro-ph},
       adsurl = {https://ui.adsabs.harvard.edu/abs/2007MNRAS.381...41H}
}

@ARTICLE{1982ApJ...255..382H,
       author = {{Hickson}, P.},
        title = "{Systematic properties of compact groups of galaxies.}",
      journal = {\apj},
         year = 1982,
        month = apr,
       volume = {255},
        pages = {382-391},
          doi = {10.1086/159838},
       adsurl = {https://ui.adsabs.harvard.edu/abs/1982ApJ...255..382H}
}

@ARTICLE{1992ApJ...399..353H,
       author = {{Hickson}, Paul and {Mendes de Oliveira}, Claudia and {Huchra}, John P. and {Palumbo}, Giorgio G.},
        title = "{Dynamical Properties of Compact Groups of Galaxies}",
      journal = {\apj},
         year = 1992,
        month = nov,
       volume = {399},
        pages = {353},
          doi = {10.1086/171932},
       adsurl = {https://ui.adsabs.harvard.edu/abs/1992ApJ...399..353H}
}

@Article{Hunter:2007,
  Author    = {Hunter, J. D.},
  Title     = {Matplotlib: A 2D graphics environment},
  Journal   = {Computing In Science \& Engineering},
  Volume    = {9},
  Number    = {3},
  Pages     = {90--95},
  publisher = {IEEE COMPUTER SOC},
  year      = 2007
}

@ARTICLE{2002ApJ...566..641H,
       author = {{Hoyle}, Fiona and {Vogeley}, Michael S.},
        title = "{Voids in the Point Source Catalogue Survey and the Updated Zwicky Catalog}",
      journal = {\apj},
         year = 2002,
        month = feb,
       volume = {566},
       number = {2},
        pages = {641-651},
          doi = {10.1086/338340},
archivePrefix = {arXiv},
       eprint = {astro-ph/0109357},
 primaryClass = {astro-ph},
       adsurl = {https://ui.adsabs.harvard.edu/abs/2002ApJ...566..641H}
}

@ARTICLE{2012MNRAS.426.3041H,
       author = {{Hoyle}, Fiona and {Vogeley}, M.~S. and {Pan}, D.},
        title = "{Photometric properties of void galaxies in the Sloan Digital Sky Survey Data Release 7}",
      journal = {\mnras},
         year = 2012,
        month = nov,
       volume = {426},
       number = {4},
        pages = {3041-3050},
          doi = {10.1111/j.1365-2966.2012.21943.x},
archivePrefix = {arXiv},
       eprint = {1205.1843},
 primaryClass = {astro-ph.CO},
       adsurl = {https://ui.adsabs.harvard.edu/abs/2012MNRAS.426.3041H}
}

@ARTICLE{2024MNRAS.527.4087J,
       author = {{Jaber}, Mariana and {Peper}, Marius and {Hellwing}, Wojciech A. and {Arag{\'o}n-Calvo}, Miguel A. and {Valenzuela}, Octavio},
        title = "{Hierarchical structure of the cosmic web and galaxy properties}",
      journal = {\mnras},
         year = 2024,
        month = jan,
       volume = {527},
       number = {2},
        pages = {4087-4099},
          doi = {10.1093/mnras/stad3347},
archivePrefix = {arXiv},
       eprint = {2304.14387},
 primaryClass = {astro-ph.GA},
       adsurl = {https://ui.adsabs.harvard.edu/abs/2024MNRAS.527.4087J}
}

@ARTICLE{2020A&A...639A..71K,
       author = {{Kuutma}, Teet and {Poudel}, Anup and {Einasto}, Maret and {Hein{\"a}m{\"a}ki}, Pekka and {Lietzen}, Heidi and {Tamm}, Antti and {Tempel}, Elmo},
        title = "{Properties of brightest group galaxies in cosmic web filaments}",
      journal = {\aap},
         year = 2020,
        month = jul,
       volume = {639},
          eid = {A71},
        pages = {A71},
          doi = {10.1051/0004-6361/201937282},
archivePrefix = {arXiv},
       eprint = {2006.04463},
 primaryClass = {astro-ph.GA},
       adsurl = {https://ui.adsabs.harvard.edu/abs/2020A&A...639A..71K}
}

@ARTICLE{2017MNRAS.470...85L,
       author = {{Lares}, Marcelo and {Luparello}, Heliana E. and {Maldonado}, Victoria and {Ruiz}, Andr{\'e}s N. and {Paz}, Dante J. and {Ceccarelli}, Laura and {Garcia Lambas}, Diego},
        title = "{Voids and superstructures: correlations and induced large-scale velocity flows}",
      journal = {\mnras},
         year = 2017,
        month = sep,
       volume = {470},
       number = {1},
        pages = {85-94},
          doi = {10.1093/mnras/stx1227},
archivePrefix = {arXiv},
       eprint = {1705.06541},
 primaryClass = {astro-ph.CO},
       adsurl = {https://ui.adsabs.harvard.edu/abs/2017MNRAS.470...85L}
}

@ARTICLE{2024MNRAS.527.2663L,
       author = {{Li}, Gang and {Ma}, Yin-Zhe and {Tramonte}, Denis and {Li}, Guo-Liang},
        title = "{Cross-correlation of cosmic voids with thermal Sunyaev-Zel'dovich data}",
      journal = {\mnras},
         year = 2024,
        month = jan,
       volume = {527},
       number = {2},
        pages = {2663-2671},
          doi = {10.1093/mnras/stad3396},
archivePrefix = {arXiv},
       eprint = {2311.00826},
 primaryClass = {astro-ph.CO},
       adsurl = {https://ui.adsabs.harvard.edu/abs/2024MNRAS.527.2663L}
}

@ARTICLE{2012A&A...545A.104L,
       author = {{Lietzen}, H. and {Tempel}, E. and {Hein{\"a}m{\"a}ki}, P. and {Nurmi}, P. and {Einasto}, M. and {Saar}, E.},
        title = "{Environments of galaxies in groups within the supercluster-void network}",
      journal = {\aap},
         year = 2012,
        month = sep,
       volume = {545},
          eid = {A104},
        pages = {A104},
          doi = {10.1051/0004-6361/201219353},
archivePrefix = {arXiv},
       eprint = {1207.7070},
 primaryClass = {astro-ph.CO},
       adsurl = {https://ui.adsabs.harvard.edu/abs/2012A&A...545A.104L}
}

@ARTICLE{2015ApJ...810..165L,
       author = {{Liu}, Chen-Xu and {Pan}, Danny C. and {Hao}, Lei and {Hoyle}, Fiona and {Constantin}, Anca and {Vogeley}, Michael S.},
        title = "{Spectral Properties of Galaxies in Void Regions}",
      journal = {\apj},
         year = 2015,
        month = sep,
       volume = {810},
       number = {2},
          eid = {165},
        pages = {165},
          doi = {10.1088/0004-637X/810/2/165},
archivePrefix = {arXiv},
       eprint = {1509.04430},
 primaryClass = {astro-ph.GA},
       adsurl = {https://ui.adsabs.harvard.edu/abs/2015ApJ...810..165L}
}

@ARTICLE{1987ApJ...321..622M,
       author = {{Mamon}, Gary A.},
        title = "{The Dynamics of Small Groups of Galaxies. I. Virialized Groups}",
      journal = {\apj},
         year = 1987,
        month = oct,
       volume = {321},
        pages = {622},
          doi = {10.1086/165658},
       adsurl = {https://ui.adsabs.harvard.edu/abs/1987ApJ...321..622M}
}

@ARTICLE{2017ApJ...835..161M,
       author = {{Mao}, Qingqing and {Berlind}, Andreas A. and {Scherrer}, Robert J. and {Neyrinck}, Mark C. and {Scoccimarro}, Rom{\'a}n and {Tinker}, Jeremy L. and {McBride}, Cameron K. and {Schneider}, Donald P. and {Pan}, Kaike and {Bizyaev}, Dmitry and {Malanushenko}, Elena and {Malanushenko}, Viktor},
        title = "{A Cosmic Void Catalog of SDSS DR12 BOSS Galaxies}",
      journal = {\apj},
         year = 2017,
        month = feb,
       volume = {835},
       number = {2},
          eid = {161},
        pages = {161},
          doi = {10.3847/1538-4357/835/2/161},
archivePrefix = {arXiv},
       eprint = {1602.02771},
 primaryClass = {astro-ph.CO},
       adsurl = {https://ui.adsabs.harvard.edu/abs/2017ApJ...835..161M}
}

@ARTICLE{2002ApJ...569..101M,
       author = {{Marinoni}, Christian and {Hudson}, Michael J.},
        title = "{The Mass-to-Light Function of Virialized Systems and the Relationship between Their Optical and X-Ray Properties}",
      journal = {\apj},
         year = 2002,
        month = apr,
       volume = {569},
       number = {1},
        pages = {101-111},
          doi = {10.1086/339319},
archivePrefix = {arXiv},
       eprint = {astro-ph/0109134},
 primaryClass = {astro-ph},
       adsurl = {https://ui.adsabs.harvard.edu/abs/2002ApJ...569..101M}
}

@ARTICLE{2020MNRAS.491.5747M,
       author = {{Martizzi}, Davide and {Vogelsberger}, Mark and {Torrey}, Paul and {Pillepich}, Annalisa and {Hansen}, Steen H. and {Marinacci}, Federico and {Hernquist}, Lars},
        title = "{Baryons in the Cosmic Web of IllustrisTNG - II. The connection among galaxies, haloes, their formation time, and their location in the Cosmic Web}",
      journal = {\mnras},
         year = 2020,
        month = feb,
       volume = {491},
       number = {4},
        pages = {5747-5758},
          doi = {10.1093/mnras/stz3418},
archivePrefix = {arXiv},
       eprint = {1907.04333},
 primaryClass = {astro-ph.GA},
       adsurl = {https://ui.adsabs.harvard.edu/abs/2020MNRAS.491.5747M}
}

@ARTICLE{2002MNRAS.335..216M,
       author = {{Merch{\'a}n}, Manuel and {Zandivarez}, Ariel},
        title = "{Galaxy groups in the 2dF Galaxy Redshift Survey: the catalogue}",
      journal = {\mnras},
         year = 2002,
        month = sep,
       volume = {335},
       number = {1},
        pages = {216-222},
          doi = {10.1046/j.1365-8711.2002.05623.x},
archivePrefix = {arXiv},
       eprint = {astro-ph/0204493},
 primaryClass = {astro-ph},
       adsurl = {https://ui.adsabs.harvard.edu/abs/2002MNRAS.335..216M}
}

@ARTICLE{2010MNRAS.409..936P,
       author = {{Padilla}, Nelson and {Lambas}, Diego Garc{\'\i}a. and {Gonz{\'a}lez}, Roberto},
        title = "{Local and global environmental effects on galaxies and active galactic nuclei}",
      journal = {\mnras},
         year = 2010,
        month = dec,
       volume = {409},
       number = {3},
        pages = {936-952},
          doi = {10.1111/j.1365-2966.2010.17396.x},
archivePrefix = {arXiv},
       eprint = {0911.5345},
 primaryClass = {astro-ph.CO},
       adsurl = {https://ui.adsabs.harvard.edu/abs/2010MNRAS.409..936P}
}

@ARTICLE{2012MNRAS.421..926P,
       author = {{Pan}, Danny C. and {Vogeley}, Michael S. and {Hoyle}, Fiona and {Choi}, Yun-Young and {Park}, Changbom},
        title = "{Cosmic voids in Sloan Digital Sky Survey Data Release 7}",
      journal = {\mnras},
         year = 2012,
        month = apr,
       volume = {421},
       number = {2},
        pages = {926-934},
          doi = {10.1111/j.1365-2966.2011.20197.x},
archivePrefix = {arXiv},
       eprint = {1103.4156},
 primaryClass = {astro-ph.CO},
       adsurl = {https://ui.adsabs.harvard.edu/abs/2012MNRAS.421..926P}
}

@ARTICLE{2009MNRAS.400.1105P,
       author = {{Park}, Daeseong and {Lee}, Jounghun},
        title = "{The bridge effect of void filaments}",
      journal = {\mnras},
         year = 2009,
        month = dec,
       volume = {400},
       number = {2},
        pages = {1105-1108},
          doi = {10.1111/j.1365-2966.2009.15524.x},
archivePrefix = {arXiv},
       eprint = {0905.4277},
 primaryClass = {astro-ph.CO},
       adsurl = {https://ui.adsabs.harvard.edu/abs/2009MNRAS.400.1105P}
}

@ARTICLE{2017MNRAS.471....2P,
       author = {{Paul}, S. and {John}, R.~S. and {Gupta}, P. and {Kumar}, H.},
        title = "{Understanding `galaxy groups' as a unique structure in the universe}",
      journal = {\mnras},
         year = 2017,
        month = oct,
       volume = {471},
       number = {1},
        pages = {2-11},
          doi = {10.1093/mnras/stx1488},
archivePrefix = {arXiv},
       eprint = {1706.01916},
 primaryClass = {astro-ph.CO},
       adsurl = {https://ui.adsabs.harvard.edu/abs/2017MNRAS.471....2P}
}

@ARTICLE{2025A&A...695A..84P,
       author = {{P{\'e}rez}, I. and {Gil}, L. and {Ferr{\'e}-Mateu}, A. and {Torres-R{\'\i}os}, G. and {Zurita}, A. and {Argudo-Fern{\'a}ndez}, M. and {Bidaran}, B. and {S{\'a}nchez-Menguiano}, L. and {Ruiz-Lara}, T. and {Dom{\'\i}nguez-G{\'o}mez}, J. and {Duarte Puertas}, S. and {Espada}, D. and {Falc{\'o}n-Barroso}, J. and {Florido}, E. and {Garc{\'\i}a-Benito}, R. and {Jim{\'e}nez}, A. and {Peletier}, R.~F. and {Rom{\'a}n}, J. and {S{\'a}nchez Alarc{\'o}n}, P. and {S{\'a}nchez-Bl{\'a}zquez}, P. and {V{\'a}squez-Bustos}, P.},
        title = "{Galaxy mass-size segregation in the cosmic web from the CAVITY parent sample}",
      journal = {\aap},
         year = 2025,
        month = mar,
       volume = {695},
          eid = {A84},
        pages = {A84},
          doi = {10.1051/0004-6361/202452514},
archivePrefix = {arXiv},
       eprint = {2501.07345},
 primaryClass = {astro-ph.GA},
       adsurl = {https://ui.adsabs.harvard.edu/abs/2025A&A...695A..84P}
}

@INPROCEEDINGS{2025hsa..conf...93P,
       author = {{P{\'e}rez}, I. and {Verley}, S. and {S{\'a}nchez-Menguiano}, L. and {Ruiz-Lara}, T. and {Garc{\'\i}a-Benito}, R. and {Duarte Puertas}, S. and {Jim{\'e}nez}, A. and {Dom{\'\i}nguez-G{\'o}mez}, J. and {Espada}, D. and {Peletier}, R.~F. and {Rom{\'a}n}, J. and {Rodr{\'\i}guez}, M.~I. and {Argudo-Fern{\'a}ndez}, M. and {Torres-R{\'\i}os}, G. and {Bidaran}, B. and {Alc{\'a}zar-Laynez}, M. and {van de Weygaert}, R. and {S{\'a}nchez}, S.~F. and {Lisenfeld}, U. and {Zurita}, A. and {Florido}, E. and {van der Hulst}, J.~M. and {Bl{\'a}zquez-Calero}, G. and {Villalba-Gonz{\'a}lez}, P. and {del Moral-Castro}, I. and {S{\'a}nchez Alarc{\'o}n}, P. and {Lugo-Aranda}, A. and {Canossa}, M. and {Conrado}, A. and {De Daniloff}, S.~B. and {Gonz{\'a}lez Delgado}, R. and {Falc{\'o}n-Barroso}, J. and {Ferr{\'e}-Mateu}, A. and {Hern{\'a}ndez-S{\'a}nchez}, M. and {Awad}, P. and {Kreckel}, K. and {Courtois}, H. and {Galbany}, L. and {S{\'a}nchez-Bl{\'a}zquez}, P. and {P{\'e}rez-Montero}, E. and {S{\'a}nchez-Portal}, M. and {Bongiovanni}, A. and {Planelles}, S. and {Quilis}, V. and {Vasquez-Bustos}, P. and {Weijmans}, A. and {Raj}, M.~A. and {Arag{\'o}n-Calvo}, M.~A. and {Ag{\"u}{\'\i}-Fern{\'a}ndez}, J.~F. and {Bla{\v{z}}ek}, M. and {Bergond}, G. and {Fern{\'a}ndez-Mart{\'\i}n}, A. and {Flores}, J. and {G{\'o}ngora}, S. and {Guijarro}, A. and {Hermelo}, I. and {Pinter}, V. and {Vico Linares}, J.~I.},
        title = "{Exploring Galaxies in Cosmic Voids: CAVITY and its First Public Data Release (DR1)}",
    booktitle = {Highlights of Spanish Astrophysics XII},
         year = 2025,
       editor = {{Manteiga}, M. and {Gonz{\'a}lez-Galindo}, F. and {Labiano-Ortega}, A. and {Mart{\'\i}nez-Gonz{\'a}lez}, M.~J. and {Rea}, N. and {Romero-G{\'o}mez}, M. and {Ulla-Miguel}, A. and {Yepes}, G. and {Rodr{\'\i}guez-L{\'o}pez}, C. and {G{\'o}mez-Garc{\'\i}a}, A. and {Dafonte}, C.},
        month = may,
        pages = {93},
       adsurl = {https://ui.adsabs.harvard.edu/abs/2025hsa..conf...93P}
}

@ARTICLE{2024A&A...689A.213P,
       author = {{P{\'e}rez}, I. and {Verley}, S. and {S{\'a}nchez-Menguiano}, L. and {Ruiz-Lara}, T. and {Garc{\'\i}a-Benito}, R. and {Duarte Puertas}, S. and {Jim{\'e}nez}, A. and {Dom{\'\i}nguez-G{\'o}mez}, J. and {Espada}, D. and {Peletier}, R.~F. and {Rom{\'a}n}, J. and {Rodr{\'\i}guez}, M.~I. and {Argudo-Fern{\'a}ndez}, M. and {Torres-R{\'\i}os}, G. and {Bidaran}, B. and {Alc{\'a}zar-Laynez}, M. and {van de Weygaert}, R. and {S{\'a}nchez}, S.~F. and {Lisenfeld}, U. and {Zurita}, A. and {Florido}, E. and {van der Hulst}, J.~M. and {Bl{\'a}zquez-Calero}, G. and {Villalba-Gonz{\'a}lez}, P. and {del Moral-Castro}, I. and {S{\'a}nchez Alarc{\'o}n}, P. and {Lugo-Aranda}, A. and {Walo-Mart{\'\i}n}, D. and {Conrado}, A. and {Gonz{\'a}lez Delgado}, R. and {Falc{\'o}n-Barroso}, J. and {Ferr{\'e}-Mateu}, A. and {Hern{\'a}ndez-S{\'a}nchez}, M. and {Awad}, P. and {Kreckel}, K. and {Courtois}, H. and {Espada-Miura}, R. and {Rela{\~n}o}, M. and {Galbany}, L. and {S{\'a}nchez-Bl{\'a}zquez}, P. and {P{\'e}rez-Montero}, E. and {S{\'a}nchez-Portal}, M. and {Bongiovanni}, A. and {Planelles}, S. and {Quilis}, V. and {Weijmans}, A. and {Raj}, M.~A. and {Arag{\'o}n-Calvo}, M.~A. and {Azzaro}, M. and {Bergond}, G. and {Blazek}, M. and {Cikota}, S. and {Fern{\'a}ndez-Mart{\'\i}n}, A. and {Gardini}, A. and {Guijarro}, A. and {Hermelo}, I. and {Mart{\'\i}n}, P. and {Vico Linares}, J.~I.},
        title = "{CAVITY, Calar Alto Void Integral-field Treasury surveY and project extension}",
      journal = {\aap},
         year = 2024,
        month = sep,
       volume = {689},
          eid = {A213},
        pages = {A213},
          doi = {10.1051/0004-6361/202449749},
archivePrefix = {arXiv},
       eprint = {2405.04217},
 primaryClass = {astro-ph.GA},
       adsurl = {https://ui.adsabs.harvard.edu/abs/2024A&A...689A.213P}
}

@Article{PER-GRA:2007,
  Author    = {P\'erez, Fernando and Granger, Brian E.},
  Title     = {{IP}ython: a System for Interactive Scientific Computing},
  Journal   = {Computing in Science and Engineering},
  Volume    = {9},
  Number    = {3},
  Pages     = {21--29},
  month     = may,
  year      = 2007,
  url       = "https://ipython.org",
  ISSN      = "1521-9615",
  doi       = {10.1109/MCSE.2007.53},
  publisher = {IEEE Computer Society},
}

@ARTICLE{2003AJ....126.1677P,
       author = {{Pisani}, Armando and {Ramella}, Massimo and {Geller}, Margaret J.},
        title = "{The Mass Function and Distribution of Velocity Dispersions for UZC Groups of Galaxies}",
      journal = {\aj},
         year = 2003,
        month = oct,
       volume = {126},
       number = {4},
        pages = {1677-1689},
          doi = {10.1086/377621},
archivePrefix = {arXiv},
       eprint = {astro-ph/0308133},
 primaryClass = {astro-ph},
       adsurl = {https://ui.adsabs.harvard.edu/abs/2003AJ....126.1677P}
}

@ARTICLE{2007A&A...464..451P,
       author = {{Popesso}, P. and {Biviano}, A. and {B{\"o}hringer}, H. and {Romaniello}, M.},
        title = "{RASS-SDSS galaxy cluster survey. VII. On the cluster mass-to-light ratio and the halo occupation distribution}",
      journal = {\aap},
         year = 2007,
        month = mar,
       volume = {464},
       number = {2},
        pages = {451-464},
          doi = {10.1051/0004-6361:20054708},
archivePrefix = {arXiv},
       eprint = {astro-ph/0606260},
 primaryClass = {astro-ph},
       adsurl = {https://ui.adsabs.harvard.edu/abs/2007A&A...464..451P}
}

@ARTICLE{2023MNRAS.524.5768P,
       author = {{Porter}, Lori E. and {Holwerda}, Benne W. and {Kruk}, Sandor and {Lara-L{\'o}pez}, Maritza and {Pimbblet}, Kevin A. and {Henry}, Christopher P.~A. and {Casura}, Sarah and {Kelvin}, Lee S.},
        title = "{The loneliest galaxies in the Universe: a GAMA and Galaxy Zoo study on void galaxy morphology}",
      journal = {\mnras},
         year = 2023,
        month = oct,
       volume = {524},
       number = {4},
        pages = {5768-5780},
          doi = {10.1093/mnras/stad1125},
archivePrefix = {arXiv},
       eprint = {2304.05999},
 primaryClass = {astro-ph.GA},
       adsurl = {https://ui.adsabs.harvard.edu/abs/2023MNRAS.524.5768P}
}

@ARTICLE{2002A&A...389..405P,
       author = {{Pustilnik}, S.~A. and {Martin}, J. -M. and {Huchtmeier}, W.~K. and {Brosch}, N. and {Lipovetsky}, V.~A. and {Richter}, G.~M.},
        title = "{Studies of galaxies in voids. I. H I observations of Blue Compact Galaxies}",
      journal = {\aap},
         year = 2002,
        month = jul,
       volume = {389},
        pages = {405-418},
          doi = {10.1051/0004-6361:20020591},
archivePrefix = {arXiv},
       eprint = {astro-ph/0205195},
 primaryClass = {astro-ph},
       adsurl = {https://ui.adsabs.harvard.edu/abs/2002A&A...389..405P}
}

@ARTICLE{2014MNRAS.445.4045R,
       author = {{Ricciardelli}, E. and {Cava}, A. and {Varela}, J. and {Quilis}, V.},
        title = "{The star formation activity in cosmic voids}",
      journal = {\mnras},
         year = 2014,
        month = dec,
       volume = {445},
       number = {4},
        pages = {4045-4054},
          doi = {10.1093/mnras/stu2061},
archivePrefix = {arXiv},
       eprint = {1410.0023},
 primaryClass = {astro-ph.GA},
       adsurl = {https://ui.adsabs.harvard.edu/abs/2014MNRAS.445.4045R}
}

@ARTICLE{2013MNRAS.435..222R,
       author = {{Rieder}, Steven and {van de Weygaert}, Rien and {Cautun}, Marius and {Beygu}, Burcu and {Portegies Zwart}, Simon},
        title = "{Assembly of filamentary void galaxy configurations}",
      journal = {\mnras},
         year = 2013,
        month = oct,
       volume = {435},
       number = {1},
        pages = {222-241},
          doi = {10.1093/mnras/stt1288},
archivePrefix = {arXiv},
       eprint = {1307.7182},
 primaryClass = {astro-ph.CO},
       adsurl = {https://ui.adsabs.harvard.edu/abs/2013MNRAS.435..222R}
}

@ARTICLE{2024MNRAS.528.2822R,
       author = {{Rodr{\'\i}guez-Medrano}, Agust{\'\i}n M. and {Springel}, Volker and {Stasyszyn}, Federico A. and {Paz}, Dante J.},
        title = "{The evolutionary path of void galaxies in TNG300 simulation}",
      journal = {\mnras},
         year = 2024,
        month = feb,
       volume = {528},
       number = {2},
        pages = {2822-2833},
          doi = {10.1093/mnras/stae193},
archivePrefix = {arXiv},
       eprint = {2312.09297},
 primaryClass = {astro-ph.GA},
       adsurl = {https://ui.adsabs.harvard.edu/abs/2024MNRAS.528.2822R}
}

@ARTICLE{2023MNRAS.521..916R,
       author = {{Rodr{\'\i}guez-Medrano}, Agust{\'\i}n M. and {Paz}, Dante J. and {Stasyszyn}, Federico A. and {Rodr{\'\i}guez}, Facundo and {Ruiz}, Andr{\'e}s N. and {Merch{\'a}n}, Manuel},
        title = "{Local and large-scale effects on the astrophysics of void galaxies}",
      journal = {\mnras},
         year = 2023,
        month = may,
       volume = {521},
       number = {1},
        pages = {916-925},
          doi = {10.1093/mnras/stad623},
archivePrefix = {arXiv},
       eprint = {2212.10594},
 primaryClass = {astro-ph.GA},
       adsurl = {https://ui.adsabs.harvard.edu/abs/2023MNRAS.521..916R}
}

@ARTICLE{2004ApJ...617...50R,
       author = {{Rojas}, Randall R. and {Vogeley}, Michael S. and {Hoyle}, Fiona and {Brinkmann}, Jon},
        title = "{Photometric Properties of Void Galaxies in the Sloan Digital Sky Survey}",
      journal = {\apj},
         year = 2004,
        month = dec,
       volume = {617},
       number = {1},
        pages = {50-63},
          doi = {10.1086/425225},
archivePrefix = {arXiv},
       eprint = {astro-ph/0307274},
 primaryClass = {astro-ph},
       adsurl = {https://ui.adsabs.harvard.edu/abs/2004ApJ...617...50R}
}

@ARTICLE{2022MNRAS.517..712R,
       author = {{Rosas-Guevara}, Yetli and {Tissera}, Patricia and {Lagos}, Claudia del P. and {Paillas}, Enrique and {Padilla}, Nelson},
        title = "{Revealing the properties of void galaxies and their assembly using the EAGLE simulation}",
      journal = {\mnras},
         year = 2022,
        month = nov,
       volume = {517},
       number = {1},
        pages = {712-731},
          doi = {10.1093/mnras/stac2583},
archivePrefix = {arXiv},
       eprint = {2204.04565},
 primaryClass = {astro-ph.GA},
       adsurl = {https://ui.adsabs.harvard.edu/abs/2022MNRAS.517..712R}
}

@software{reback2020pandas,
    author       = {The pandas development team},
    title        = {pandas-dev/pandas: Pandas},
    month        = feb,
    year         = 2020,
    publisher    = {Zenodo},
    version      = {latest},
    doi          = {10.5281/zenodo.3509134},
    url          = {https://doi.org/10.5281/zenodo.3509134}
}





\end{document}